\def\mearth{{\rm\,M_\oplus}}

\def\deg{^\circ}

\documentclass[preprint,authoryear,12pt]{elsarticle}
\usepackage{amssymb}
\usepackage{aas_macros}

\begin{document}
 
\begin{frontmatter}
\title{Origin and dynamical evolution of the asteroid belt}

\author{Sean N. Raymond$^{a}$ \& David Nesvorn\'y$^{b}$\\
\address[1]{Laboratoire d'Astrophysique de Bordeaux, Univ. Bordeaux, CNRS, B18N, All{\'e}e Geoffroy Saint-Hilaire, 33615 Pessac, France; rayray.sean@gmail.com}
\address[2]{Department of Space Studies, Southwest Research Institute, 1050 Walnut St., Suite 300, Boulder, CO 80302, USA; davidn@boulder.swri.edu}}

\begin{abstract}
The asteroid belt was dynamically shaped during and after planet formation. Despite representing a broad ring of stable orbits, the belt contains less than one one-thousandth of an Earth mass. The asteroid orbits are dynamically excited with a wide range in eccentricity and inclination and their compositions are diverse, with a general trend toward dry objects in the inner belt and more water-rich objects in the outer belt. Here we review models of the asteroid belt's origins and dynamical history. The classical view is that the belt was born with several Earth masses in planetesimals, then strongly depleted. However, it is possible that very few planetesimals ever formed in the asteroid region and that the belt's story is one of implantation rather than depletion. A number of processes may have implanted asteroids from different regions of the Solar System, dynamically removed them, and excited their orbits. During the gaseous disk phase these include the effects of giant planet growth and migration and sweeping secular resonances. After the gaseous disk phase these include scattering from resident planetary embryos, chaos in the giant planets' orbits, the giant planet instability, and long-term dynamical evolution. Different global models for Solar System formation imply contrasting dynamical histories of the asteroid belt. Vesta and Ceres may have been implanted from opposite regions of the Solar System -- Ceres from the Jupiter-Saturn region and Vesta from the terrestrial planet region -- and could therefore represent very different formation conditions. 
\end{abstract}
\end{frontmatter}

\footnotesize
\noindent {\tt Chapter to appear in {\em Vesta and Ceres: Insights into the Dawn of the Solar System} (Editors Simone Marchi, Carol A. Raymond, and Christopher T. Russell; Cambridge University Press)}
\normalsize

\section{The asteroid belt in the context of Solar System formation} 

The asteroid belt marks the boundary between the rocky and gaseous planets. It is the widest piece of Solar System real estate between Mercury and Neptune that does not contain a planet (as measured in dynamical terms: the orbital period ratio between inner and outer stable orbits within the belt is $>2$, wider than the dynamical spacing between many planets). The belt's total mass is less than one one-thousandth of an Earth mass~\citep{krasinsky02,kuchynka13,demeo13}. Asteroids' orbits are dynamically excited, with eccentricities ranging from zero to above 0.3 and inclinations from zero to $>20\deg$. The belt's composition varies radially, with S-class objects most common in the inner main belt, C-class objects dominating the outer main and substantial mixing between the two populations~\citep{gradie82,demeo13,demeo14}. S-types are spectroscopically linked~\citep{burbine02} with ordinary chondrite meteorites, which have little water~\citep{robert77,alexander18} and C-types with carbonaceous chondrites, which typically have $\sim 10\%$ water by mass~\citep{kerridge85,alexander18}.

Understanding the asteroid belt's dynamical history is challenging. In general the more mass in a dynamical system, the more excited it becomes by self-stirring. For instance, a higher-mass disk of planetesimals excites itself due to mutual gravitational interactions faster and to a higher degree than a lower-mass disk~\citep[see, for example, ][]{wetherill89,ida92,kokubo00}. The excitation is driven mainly by the most massive bodies, which grow faster and more massive in higher-mass disks.

Models for the asteroid belt's origins make assumptions about the belt's birth conditions. The classical view is that the belt was born with several Earth masses in planetesimals and was subsequently depleted by a factor of $\sim 1000$. That view is a legacy of a terrestrial planet formation that assumes that disks follow smooth radial surface density profiles~\citep{wetherill80,chambers98,raymond14}. An alternate view suggests that the belt was born with very little mass -- perhaps none at all -- and was later populated by planetesimals that originated in other parts of the Sun's planet-forming disk~\citep[e.g.][]{hansen09,raymond17b}. Such an assumption is motivated by observations of ringed substructure seen in the dust component of protoplanetary disks~\citep[e.g.][]{alma15,andrews16}, as well as models of planetesimal formation in localized concentrations of dust and pebbles~\citep[e.g.][]{johansen14,birnstiel16}.  

Here we review the dynamical processes that may have sculpted the asteroid belt over its history. In the next two subsections we describe the observed structure of the asteroid belt and present a rough Solar System timeline as inferred from cosmochemical constraints (see also Chapter 13).  In Sect. 2 we examine dynamical processes that may have affected the asteroid belt during the gaseous disk phase.  In Sect. 3 we describe the key processes affecting the belt after gas disk dispersal.  In Sect. 4 we show how the asteroid belt's story fits within global models of Solar System formation, discuss the origins of Ceres and Vesta in the context of different models, and highlight outstanding problems.

\subsection{The asteroid belt's observed structure} 
\begin{figure}
\begin{center}
\includegraphics[width=1.06\textwidth]{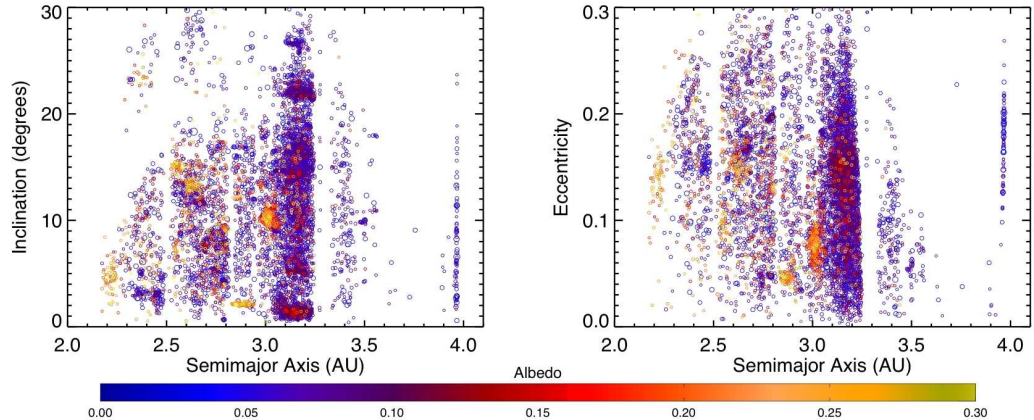} 
    \caption[]{Orbital and compositional distribution of 8237 asteroids larger than 10~km in diameter. The symbol size is proportional to actual size and the colors represent asteroid albedos. Typical albedos of C-types are $\sim 5$\% and of S-types $\sim 20$\%~\citep{pravec12}.  Asteroids with albedos above 0.3 were given colors corresponding to albedos of 0.3. Asteroid data from mp3c.oca.eu.}
         \label{fig:mainbelt}
         \end{center}
\end{figure}

Figure~\ref{fig:mainbelt} shows the orbital distribution of asteroids with diameters $D>10$~km. We can distill the asteroid belt's characteristics down to four broad constraints that must be matched by any model of Solar System evolution. These are:
\begin{enumerate}
\item The belt's very low total mass~\citep[$\sim 4.5 \times 10^{-4} \mearth$;][]{krasinsky02,kuchynka13,demeo13}.  Most of the mass is contained within just four asteroids: Ceres (31\% of the total mass of the belt), Vesta (9\%), Pallas (7\%) and Hygiea (3\%). 
\item The orbital structure of the belt.  Asteroid orbits are excited, with eccentricities $0 < e < 0.3$ and inclinations $0 < i < 20^\circ$ (Fig.~\ref{fig:mainbelt}) and modest variations across the main belt~\citep[see Fig. 1 from ][]{morby15b}.  There is additional substructure within the belt such as the Kirkwood gaps at the 3:1, 5:3 and 2:1 resonances and clumps associated with asteroid families~\citep[see][for a review]{nesvorny15d}.
\item The compositional structure of the belt. Taxonomic classification show that the inner main belt is dominated by S-types and the outer main belt by C-types~\citep{gradie82,bus02,demeo13,demeo14}.  D-types are prevalent in the outer main belt and Jupiter's 1:1 resonant Trojan swarms~\citep{emery15}.
\item The asteroid size-frequency distribution (SFD) is top-heavy with the bulk of the belt's mass contained in the most massive objects.  Given that the asteroid SFD is likely a signature of accretion via the streaming instability~\citep[e.g.][]{youdin05,morby09,johansen15,nesvorny19} and subsequent sculpting by collisional processes rather than dynamical ones~\citep[e.g.][]{bottke05}, we will not discuss it further in this review.
\end{enumerate}

\subsection{A rough Solar System timeline}

The timeline of events that took place during Solar System formation can be roughly inferred from a combination of astrophysical and cosmochemical constraints~\citep[see also Fig. 9 of ][ and their associated discussion]{nittler16}. Here is a simplified summary of the current thinking:

\begin{itemize}

\item The inner Solar System's first solids -- Calcium-Aluminum rich Inclusions, or CAIs -- have a narrow range of formation ages centered on 4.568 Gyr~\citep[e.g.][]{bouvier10,connelly12}. This sets the Solar System's {\em time zero}, the start of planet formation.

\item Planetesimals likely formed over roughly the subsequent five million years, during the gaseous disk phase. Analysis of iron meteorites indicates that their parent bodies formed within $\sim 1$~Myr of CAIs~\citep{kruijer14,schiller15,kruijer20}, meaning that planetesimal formation was underway very quickly. Current thinking~\citep[see reviews by][]{johansen14,johansen15,birnstiel16} suggests that planetesimals formed as drifting dust was locally concentrated to a sufficient degree to trigger gas-particle instabilities such as the streaming instability~\citep{youdin05,johansen07,squire18}.  The resulting planetesimals had sizes extending to hundreds of km in diameter, with a characteristic size of $\sim$100 km~\citep{johansen15,simon17,schaefer17,abod19}.  Indeed, the bump in the size distribution of asteroids at $D \simeq 100$ km can be interpreted as a sign of efficient growth of 100-km-class asteroids (\cite{morby09}; but see \cite{weidenschilling11}).  Planetesimals that formed within 1-2 Myr of CAIs contained a high enough concentration of the active short-lived radionuclide $^{26}$Al (half life of 717,000 years) to be strongly heated and dried out~\citep{grimm93,lichtenberg16,monteux18}. These are the parent bodies of iron meteorites.

\item Mars' growth was nearly complete during the gaseous disk phase. Isotopic (Hf/W/Th) analyses of Martian meteorites indicate that it was mostly formed within a few Myr~\citep{nimmo07,dauphas11}, although a new result indicates that this timeline may be less well constrained~\citep{marchi20}. The giant planet $\sim 10 \mearth$ cores may also have formed within 1-2 Myr. Circumstantial evidence supporting this idea comes from the overlap in ages of non-carbonaceous and carbonaceous chondrite meteorites~\citep[][; see also Chapter 13]{budde16,kruijer17,kruijer20}. As the components of these meteorites -- chondrules -- are at the size scale to drift rapidly within the disk, something must kept them spatially separated. A candidate mechanism is the pressure trap generated exterior to the orbit of a large core~\citep{lambrechts14,bitsch18}. Another possibility is that these two chondrule reservoirs were kept apart by structure within the disk itself, unrelated to the formation of giant planet cores~\citep{brasser20}.

\item The gaseous disk dissipated within roughly 5~Myr. Studies of disk frequencies around star clusters of different ages indicate that disks usually last a few Myr~\citep{haisch01,mamajek09,pfalzner14}. The latest-forming chondrites (CB chondrites) formed $\sim 5$ Myr after CAIs~\citep{krot05,johnson16}, possibly indicating that is when the Sun's gas disk dissipated. If true, this would set an upper limit on the timescale of the formation of gas giants.

\item Hf/W analyses suggest that the Earth's final giant collision -- the Moon-forming impact -- happened 50-100 Myr after CAIs~\citep{kleine09}. This timeframe is consistent with estimates using different methods, e.g, by calibrating Earth's late veneer to N-body simulations~\citep{jacobson14} and from a spike in the distribution of shock degassing ages of meteorites, presumed to originate from collisions between asteroids and debris from the impact~\citep{bottke15}.  The constraint on the late veneer's total mass inferred from highly siderophile elements~\citep{day07,walker09} disappears if the magma ocean phase lasted for a sufficient time~\citep[see ][]{rubie16,morby18}.

\item The giant planets underwent a dynamical instability no later than 100 Myr after CAIs.  The so-called {\em Late Heavy Bombardment} (LHB) hypothesis was based on an inferred spike in the age distribution of craters on the Moon that implied a large flux of impactors in the inner Solar System roughly 500 Myr after CAIs~\citep{tera74,bottke17}.  The {\em Nice model} invokes an instability in the orbits of the giant planets to explain the LHB~\citep{gomes05,levison10,deienno17}.  However, new analyses of crater ages -- coupled with cosmochemical constraints and modeling -- suggest that the bombardment flux more likely simply followed a smooth decay~\citep{boehnke16,zellner17,morby18,hartmann19}. The giant planet instability -- which is needed to explain the giant planets' orbits as well as the orbital distribution of irregular satellites, the giant planets' Trojan asteroids and the Kuiper belt~\citep[see][for a review]{nesvorny18}  -- could have happened anytime within the first 100 Myr~\citep{nesvorny18b,mojzsis19}, perhaps during terrestrial planet formation~\citep{clement18}.
\end{itemize}

\section{Asteroid belt evolution during the gaseous disk phase} 

The asteroid belt's story starts with the first planetesimals. Simulations of planetesimal formation by the streaming instability find that planetesimals form with most of the mass in the largest objects, which are generally hundreds of km in size~\citep{johansen15,simon17,abod19}. The largest planetesimals grow by accreting other planetesimals~\citep[e.g.][]{greenberg78,wetherill89,kokubo98} as well as mm- to cm-sized `pebbles' that continually drift inward through the disk due to aerodynamic gas drag~\citep[e.g.][]{weidenschilling77b,ormel10,johansen17}. The varying strength of gas drag can also cause size-sorting of planetesimals.  The smallest planetesimals undergo radial drift on timescales shorter than the disk's lifetime but planetesimals larger than roughly 10~km do not lose enough energy from gas drag to significantly alter their orbital radii~\citep[although this depends on the disk properties; ][]{adachi76}. 

A massive population of planetesimals undergoes gravitational self-stirring, which acts to excite their orbital eccentricities and inclinations~\citep[e.g.][]{ida92}. While the level of excitation is low, the collisional cross sections of the largest planetesimals are augmented by gravitational focusing and these objects can undergo runaway growth~\citep{greenberg78,wetherill89,wetherill93,kokubo98,rafikov03}.  However, when the largest objects become massive enough they excite the random velocities of planetesimals to such a degree that gravitational focusing is shut down and their growth slows and transitions to the so-called {\em oligarchic} regime~\citep{ida93,kokubo98,kokubo00,leinhardt05}. The most massive objects are typically Moon- to Mars-mass (in the inner Solar System) and are called {\em planetary embryos}.

A well-studied model proposes that the asteroids' eccentricities and inclinations were excited by planetary embryos that formed within the belt~\citep{wetherill92,chambers01b,petit01,obrien07}.  For historical reasons, and because of the connection between Mars' accretion and the asteroid belt's excitation~\citep{izidoro15c}, we will discuss this theory in the context of gas-free dynamics in Sect. 3.1.

\subsection{Implantation of planetesimals driven by the giant planets' growth}

Two gas giants formed next door to the asteroid belt.  Let us consider how this process affected the belt.

The prevailing model for gas giant formation proposes that giant planets first grow cores of $\sim 5-10 \mearth$, which gravitationally accrete gas from the disk~\citep{mizuno80,pollack96,ida04,piso14}.  The early phase of gas accretion is slow and limited by cooling and contraction of the gaseous envelope~\citep{ikoma00,hubickyj05} as well as the gas supply from the disk~\citep{fung14,lambrechts17,lambrechts19b}.  Most growing giant planets may not get past this phase, which would explain why Neptunes are much more common than Jupiters among extra-solar planets~\citep[e.g.][]{suzuki16b}.  When the mass in the planet's gaseous envelope becomes comparable to the core mass, gas accretion accelerates~\citep{pollack96,ikoma01,lissauer09}. The growing planet also carves an annular gap in the gas disk~\citep{lin86,bryden99,crida06}.

\begin{figure}
\begin{center}
\includegraphics[width=0.8\textwidth]{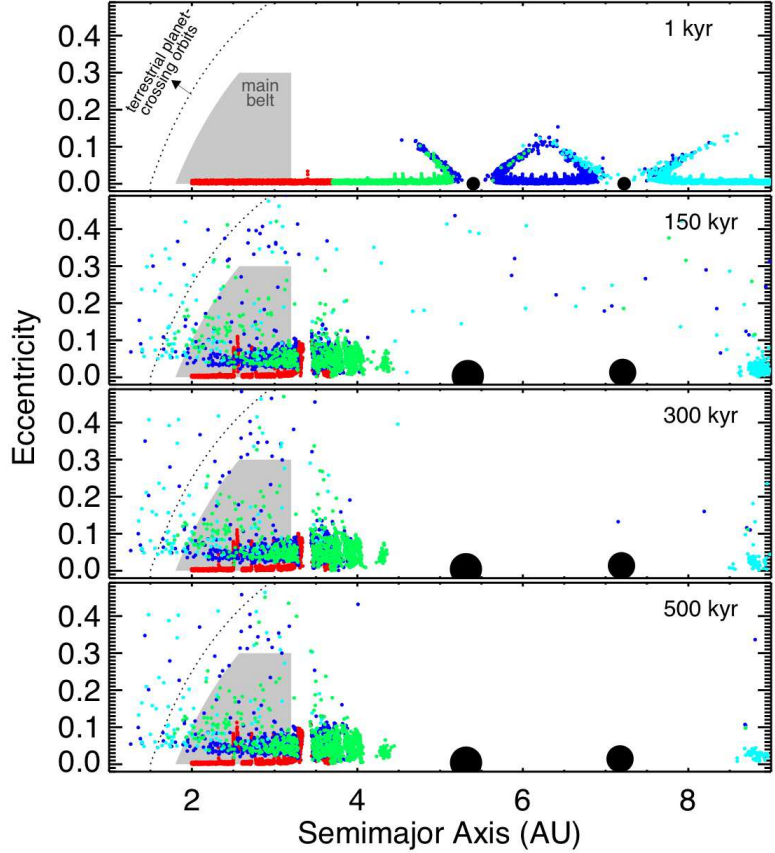} 
    \caption[]{Injection of planetesimals into the inner Solar System due to Jupiter and Saturn's rapid gas accretion~\citep[adapted from][]{raymond17}. In this simulation the giant planets' cores (initially $3 \mearth$) underwent rapid gas accretion from 100-200 kyr (for Jupiter) and 300-400 kyr (for Saturn). The gaseous disk structure responded to the giant planets' growth~\citep[from hydrodynamical models of][]{morby07a}. As the giant planets increased in mass, the orbits of nearby planetesimals (whose colors correspond to their starting orbital radii) were destabilized and gravitationally scattered by the giant planets onto eccentric orbits.  Under the action of gas drag (assuming $D=100$~km), some planetesimals were trapped on stable orbits within the main belt, originating in a region extending roughly from 4-9 AU. These may be the present-day C-type asteroids.  In this simulation giant planet migration was not accounted for, so the 5 au-wide source region for C-types is the narrowest it could possibly be; when migration is included planetesimals can be implanted into the belt from as far out as 20-30 au~\citep{raymond17,pirani19}. }
         \label{fig:injection}
         \end{center}
\end{figure}

The phase of rapid gas accretion onto a giant planet core destabilizes nearby orbits.  A planetesimal orbiting the Sun in the vicinity of a giant planet core is stable as long as its orbit remains well-separated from that of the core. The critical distance for stability of a small object orbiting the Sun near a massive core is $2 \sqrt{3} \approx 3.5$ times the core's Hill radius $R_H = a (m/3 M_\odot)^{1/3}$, where $a$ is the core's orbital radius, $m$ is its mass and $M_\odot$ is the Sun's mass~\citep{marchal82,gladman93}.  As the core rapidly increases in mass, its Hill sphere increases in size.  This means that nearby objects that were once on stable orbits are destabilized.

Figure~\ref{fig:injection} shows how Jupiter and Saturn's growth invariably implants C-types into the asteroid belt. This is a simple consequence of the combined effects of orbital destabilization, gravitational scattering and aerodynamic gas drag. There are two populations of implanted objects: those trapped on stable orbits within the asteroid belt and those scattered {\em past} the belt toward the terrestrial planets~\citep[perhaps to deliver water to the growing Earth; see][]{meech19}. The balance between these two outcomes depends on the strength of gas drag~\citep{raymond17}.  For stronger gas drag implantation is favored, and for weaker gas drag scattering toward the terrestrial planets dominates. The strength of gas drag is a function of both 1) the density of the gaseous disk and therefore the timing of scattering, as the gas disk's density decays in time; and 2) the planetesimal size, with smaller planetesimals feeling stronger drag. 

Planetesimals are implanted throughout the asteroid belt but preferentially in the outer main belt~\citep{raymond17,kretke17,ronnet18}. Implanted planetesimals roughly match the distribution of C-type asteroids. The source region for C-types implanted in the belt in the simulation from Fig.~\ref{fig:injection} was $\sim 5$~au-wide, from 4 to 9 au.  That represents an absolute minimum width for the source region because migration was not included.  When migration of Jupiter and Saturn is accounted for, the source region widens to $\sim$10-20 au~\citep[][this will be discussed in more detail in Sect. 2.2]{walsh11,raymond17,pirani19}.  This broad source region, as well as the multiple different epochs of implantation (corresponding to, at a minimum, the rapid gas accretion of Jupiter and Saturn and the inward migration of the ice giants), may help to explain the compositional diversity seen in carbonaceous chondrites and among C-types~\citep[e.g.][]{vernazza17,alexander18}.  And if the C-types and Earth's water share a common dynamical source then this naturally explains the isotopic (e.g., D/H and $^{15/14}$N) match between carbonaceous chondrites and Earth's water~\citep{marty06,marty12,meech19}. 

The results of Fig.~\ref{fig:injection} imply that a large fraction of the asteroid belt -- perhaps all of the C-types and even other classes in the outer belt -- was implanted from the Jupiter-Saturn region and beyond. This includes Ceres, the largest asteroid ($D =$~946 km), which is located at the (admittedly broad) peak in the distribution of $D=$1000~km planetesimals implanted in the simulations of \cite{raymond17}.  The idea that Ceres was implanted from the Jupiter-Saturn region appears consistent with its inferred composition, which is broadly similar to carbonaceous chondrites~\citep[e.g.][]{bland16,prettyman17,mcsween18,marchi19}. 

This process cannot completely explain the C-types. The implanted objects nicely match the radial distribution of C-types, but the model does not match their observed eccentricity and inclination distributions, and so another source of excitation is required.  In addition, the population of planetesimals that existed in the Jupiter-Saturn region when they underwent rapid gas accretion is uncertain.  Given the high efficiency of implantation~\citep{raymond17}, it is possible that they did not represent a large mass.  For example, in the simulation from Fig.~\ref{fig:injection} about 10\% of 100~km planetesimals from the Jupiter-Saturn region were implanted onto stable orbits within the main belt.  Allowing for an order of magnitude depletion, that would imply $< 0.01 \mearth$ in planetesimals in the source region. Such a low mass in planetesimals would be consistent with giant planet cores growing by pebble accretion~\citep{lambrechts12,levison15} but not by planetesimal accretion.  Finally, Jupiter's growth may also have affected the collisional environment within the belt, by increasing the typical collision velocity between asteroids~\citep{turrini12,turrini14}. 

\subsection{Effect of giant planet migration}

Migration is a central mechanism of planet formation~\citep{kley12,baruteau14}.  Planets with masses as low as Earth's and Mars' exchange angular momentum with the gas disk, which acts both to damp their orbital eccentricities and inclinations~\citep{papaloizou00,tanaka04} and to induce radial migration~\citep{goldreich80,ward86}.  Low-mass planets -- in the so-called type I regime -- migrate through the disk because of torques from non-axisymmetric density perturbations, whereas more massive planets open radial gaps in the disk and their migration is linked with the disk's viscous evolution in the type II regime~\citep{ward97}.  

\subsubsection{Effect of migrating giant planet cores}
The giant planet cores formed while the gas disk was still dense and likely migrated. The effect of the cores' migration on the inner Solar System has not been extensively studied because of large uncertainties. Some simulations show that the inward migration of large cores would have devastated the growing terrestrial planets~\citep{izidoro14}. Other simulations find that the absence of planets interior to Mercury could be explained if the seed of Jupiter's core formed very close to the Sun and migrated {\em outward}~\citep{raymond16}.  Both scenarios fall within the realm of possibility given the uncertainty in the underlying structure and evolution of the Sun's protoplanetary disk~\citep[see][]{morbyraymond16}.  Yet the fact that we have a system of outer giant planets rather than close-in super-Earths suggests that the migration of the giant planets' cores was limited, although the reason behind their staying in the outer Solar System remains a central problem in planet formation~\citep[see discussion in][]{raymond18d}.  A model based on migration and pebble accretion found that Jupiter's core may have originated at 20-25 au and undergone large-scale inward migration to end up at $\sim$5 au~\citep{bitsch15}. This begs the question of the fate of the solids that formed interior to 20-25 au. Another possibility is that Jupiter quickly transitioned to the slower, type II migration regime (perhaps because the disk's viscosity was very low) and acted as a barrier to the inward migration of the ice giants and Saturn's core~\citep{izidoro15b}. 

\subsubsection{The Grand Tack}

The Grand Tack model~\citep{walsh11} relies on a particular type of migration pathway for Jupiter and Saturn that comes directly from hydrodynamical simulations by \cite{masset01}. The scenario is as follows. After it opened a gap in the disk Jupiter migrated inward. Saturn accreted gas, started to open a gap and migrated inward rapidly.  Saturn caught up to Jupiter and became trapped in the outer 2:3 or 1:2 mean motion resonance with Jupiter. The two planets shared a common gap within the disk, which changed the flow of gas and the associated torque balance. The two planets subsequently migrated {\em outward}~\citep[this mechanism has been demonstrated by a number of authors -- ][]{masset01,morby07,pierens08,crida09,zhang10,pierens11,pierens14}.  Outward migration stopped as the disk dissipated or perhaps at an asymptotic radius if the disk was flared.  

\begin{figure}
\begin{center}
\includegraphics[width=0.8\textwidth]{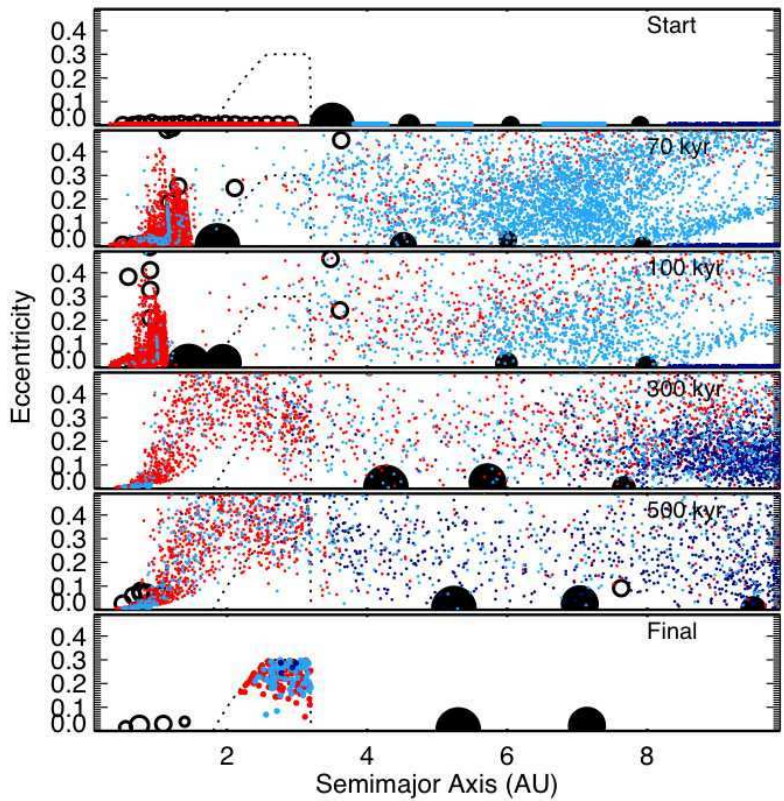} 
    \caption[]{Snapshots in the evolution of the Grand Tack model.  The simulation starts a few million years into the gaseous disk phase: Jupiter is fully grown, Saturn and the ice giants are large cores, and planetary embryos have formed in the inner Solar System.  Jupiter migrates inward through rocky planetesimals and embryos, some of which are shepherded by inner resonances and some of which are scattered outward~\citep[see][]{raymond06c,mandell07}. Saturn grows and migrates inward, is trapped in 2:3 resonance with Jupiter, and both planets migrate back outward.   Despite Jupiter's traversing the main asteroid belt is populated by S-types (red) originating interior to Jupiter's initial orbit, and C-types from belts originally located between the giant planets' cores (light blue) and from a disk of planetesimals initially exterior to the giant planets (dark blue).  The inner belt is dominated by S-types and the outer belt by C-types~\citep{walsh11,walsh12}, consistent with the compositional~\citep{gradie82,demeo14} and orbital~\citep{deienno16} structure of the present-day belt.}
%    Depletion and re-population of the asteroid belt in the Grand Tack model, adapted from \cite{walsh11}.  {\bf Panel a:} The radial distribution of ``S-type'' objects that ended up in the main belt. These objects started off between 2 and 3 au, were scattered outward by Jupiter as it migrated inward, and were then scattered back inward as Jupiter migrated back {\em outward}, ending up with roughly the same radial distribution as they started with but with an efficiency of re-implantation of $\sim$0.1\%.  {\bf Panel b:} The radial distribution of planetesimals trapped in the main belt. The red curve corresponds to S-types, which are assumed to have originated interior to Jupiter. The light and dark blue curves correspond to C-types that originated in between and beyond the giant planets, respectively. {\bf Panels c and d:} The inclination and eccentricity distributions of planetesimals trapped in the belt at the end of the simulation. Although their orbits are more excited than the actual belt, \cite{deienno16} showed that this becomes consistent with the present-day belt, given that high-eccentricity asteroids are unstable on Gyr timescales.
         \label{fig:GT}
         \end{center}
\end{figure}

The Grand Tack model invokes this inward-then-outward migration of Jupiter to sculpt the inner Solar System~\citep{walsh11,walsh12,raymond14c,jacobson14b,brasser16,walsh16,deienno19}.  The model requires that Jupiter's turnaround point was at 1.5-2 au, in which case Jupiter would deplete Mars' feeding zone but not Earth's, thus explaining the large Earth/Mars mass ratio~\citep[sometimes called the `small Mars' problem;][]{wetherill91,raymond09c}.  

In the Grand Tack scenario the asteroid belt was emptied then re-populated from multiple reservoirs (Fig.~\ref{fig:GT}). During Jupiter's inward migration, planetesimals interior to Jupiter's starting orbit were either shepherded inward or scattered outward. Planetesimals shepherded inward became building blocks of the terrestrial planets while those that were scattered outward were stranded on eccentric orbits exterior to Jupiter. When Jupiter and Saturn migrated back outward they encountered the scattered planetesimals. Most planetesimals were ejected but a fraction were scattered inward and stranded on stable orbits as the giant planets migrated away. The gas giants then encountered planetesimals that had originated farther out in the disk, between and beyond the giant planets.  Most of these were again ejected but a fraction was stranded on orbits interior to the outward-migrating giant planets.   

%The top panel of Fig.~\ref{fig:GT} shows the radial distribution of inner-disk planetesimals that started interior to Jupiter's orbit and ended up on stable orbits in the asteroid belt~\citep[from][]{walsh11}.  

S-type planetesimals that survived Jupiter's migration followed a dynamic path. Planetesimals were scattered outward then back inward, finishing on orbits similar to their initial ones. However, only $\sim 0.1\%$ of planetesimals followed this evolutionary path; most were accreted by the terrestrial planets or ejected by Jupiter. The middle panel shows how, after the giant planets' migration, the inner main belt is dominated by inner disk planetesimals and the outer main belt by planetesimals that originated between and beyond the giant planets' orbits.  This is a simple consequence of timing.  The orbital radii of scattered planetesimals correlates with Jupiter's orbital radius at the time of scattering.  Because the inner-disk planetesimals were encountered first during the giant planets' outward migration when Jupiter was closer-in, they were trapped on closer-in orbits than the outer-disk planetesimals that were encountered later when Jupiter was farther out~\citep{walsh11,walsh12}.

The bottom panel of Fig.~\ref{fig:GT} shows the orbital distribution of planetesimals trapped within the belt after the giant planets' migration was complete. While the asteroids' orbits are quite excited, they become consistent with the present-day belt as high-eccentricity ($e \gtrsim 0.2$) planetesimals are destabilized over Solar System history~\citep{deienno16}.  The Grand Tack has the strength of both exciting and depleting the belt. It can also match the belt's compositional dichotomy if one makes simple assumptions about the nature of planetesimals that formed in different regions -- that inner-disk planetesimals are linked with S-types and, as in the implantation simulation from Fig.~\ref{fig:injection}, outer disk planetesimals are linked with C-types.

Of course, the Grand Tack is built on a single migration pathway for the giant planets among many~\citep[e.g.][]{pierens14}.  While long-range outward migration of Jupiter and Saturn has been shown to be robust under certain assumptions related to the disk properties, it only holds for a range of mass ratios for Jupiter and Saturn~\citep{masset01}. It remains to be seen whether the appropriate mass ratio can be maintained during long-range outward migration when gas accretion onto the planets is taken into account in a realistic fashion.   

\subsubsection{Inward-migrating Jupiter and Saturn}

Giant planets typically migrate inward~\citep[e.g.][]{ward97}. As it migrates, the planetesimals a  giant planet encounters may be shepherded inward by mean motion resonances, accreted, or scattered~\citep[][; this is like Jupiter's inward migration in Fig.~\ref{fig:GT}]{tanaka99,fogg05,fogg07,raymond06c,mandell07,raymond16}. The balance between outcomes depends on the planet's mass and migration rate as well as the strength of the dissipation (aerodynamic or other drag) felt by the planetesimals.  For inward-migrating gas giants in current disk models, shepherding dominates.

\cite{bitsch15} showed that, including the effects of pebble accretion and migration in an evolving disk, Jupiter's core may have originated at $\sim 20$~au. \cite{pirani19} imposed this growth and migration history and studied its effect on a population of planetesimals.  They found that planetesimals were shepherded by strong resonances, as expected, and that certain populations of near-resonant asteroids (e.g., the 3:2 resonant Hildas) were broadly matched.  They also matched the observed asymmetry among Jupiter's 1:1 resonant populations (the so-called {\em Trojans}), for which the leading population outnumbers the trailing one.  In addition to specific populations of asteroids, inward migration also implants planetesimals into the outer asteroid belt, expanding the source region for C-types to be 5-20 au in width~\citep{raymond17,pirani19}, although a large fraction of implanted outer disk planetesimals originate in the present-day Jupiter-Saturn region. It remains to be seen whether this model can fully explain the Trojan population~\citep[see][]{pirani19b}.  Specific challenges include matching the Trojans' broad inclination distribution, reconciling the fact that Jupiter's Trojans are mainly D-types rather than C-types~\citep{emery15} with their source location, and that most Trojans should not survive the giant planet instability~\citep{morby05,nesvorny13}.

A purely inward migration of Jupiter would not significantly deplete the belt, so another mechanism is required to explain the belt's very low total mass.  For example, the belt may simply never have contained much mass in planetesimals; this is the foundation of the Low-mass Asteroid belt model~\citep{hansen09,drazkowska16,raymond17b}.  Global models that match all of the asteroid belt constraints will be discussed in Sect. 4~\citep[see][for building global models for Solar System- and exoplanet formation]{raymond18d}.

\subsection{Effect of secular resonance sweeping during disk dissipation} 

The gaseous disk's dissipation -- and its changing gravitational potential -- caused the locations of secular resonances with the giant planets to shift~\citep{heppenheimer80,ward81}. Certain strong resonances may have swept across the asteroid belt~\citep{lemaitre91,lecar97}. Secular resonances such as the $\nu_6$ and $\nu_{16}$ strongly excite eccentricities and inclinations of small bodies and, in the present-day Solar System, excite asteroids to the point that they become unstable and are lost from the belt~\citep[see, e.g.,][]{morby91,gladman97}.  

Sweeping secular resonances have been invoked to explain the primordial excitation of the asteroid belt~\citep{lecar97,nagasawa00,nagasawa01,nagasawa02}.  When gas drag (either aerodynamic or tidal) is accounted for, inward-sweeping secular resonances may even act to clear large planetary embryos from the belt and this process has even been invoked to explain the terrestrial planets' orbital structure~\citep{nagasawa05,thommes08c,bromley17}. However, the secular resonance sweeping model suffers from a self-consistency problem. Secular resonances require non-circular, non-coplanar giant planet orbits. The studies presented above assumed that the giant planets were on their present-day orbits during the late parts of the gaseous disk phase. But we know this was not the case because planet-disk interactions would very rapidly have decreased their orbital eccentricities and inclinations and driven the planets into a resonant configuration. This is the generic outcome of hydrodynamical simulations~\citep[e.g.][]{morby07,kley12,baruteau14,pierens14}. When plausible orbits for the giant planets are used, sweeping secular resonances are far too weak to excite the asteroids~\citep{obrien07}.  Future work may revive the sweeping secular resonance model if it is shown that giant planets can indeed maintain significant orbital eccentricities and/or inclinations during the gaseous disk phase.

\section{Evolution of the asteroid belt after dispersal of the gaseous disk}

We now turn our attention to the asteroid belt's dynamical evolution after the dispersal of the gaseous disk.  This covers the vast majority of our system's history. We first discuss gravitational self-stirring of the belt (Sect. 3.1). Next we show how asteroids are implanted from the terrestrial planet region (Sect. 3.2) and how the belt may have been chaotically excited by the giant planets (Sect. 3.3). The giant planet instability is likely to have depleted and excited the belt and implanted outer disk planetesimals (Sect. 3.4).  

\subsection{Gravitational self-stirring of the belt}
The so-called {\em classical model} of terrestrial planet formation assumes that the terrestrial planets formed locally from a massive disk of planetesimals~\citep[see][]{wetherill80,chambers01,raymond14}. There is a rich literature on the growth of rocky planetary embryos by accretion in swarms of planetesimals~\citep[e.g.][]{safronov69,greenberg78,wetherill89,kokubo00}, with a possible contribution from pebbles~\citep{levison15b,schiller18,schiller20,budde19}.  When planetary embryos enter the oligarchic regime the distribution of planetesimals' random velocities (and thus their eccentricities and inclinations) becomes dominated by gravitational scattering by embryos~\citep{ida93,kokubo00}. The level of excitation is related to the escape velocity from the largest embryos relative to the escape velocity from the Sun at the same orbital distance; this ratio is sometimes called the Safronov number~\citep{safronov69}.  

\begin{figure}
\begin{center}
\includegraphics[width=0.45\textwidth]{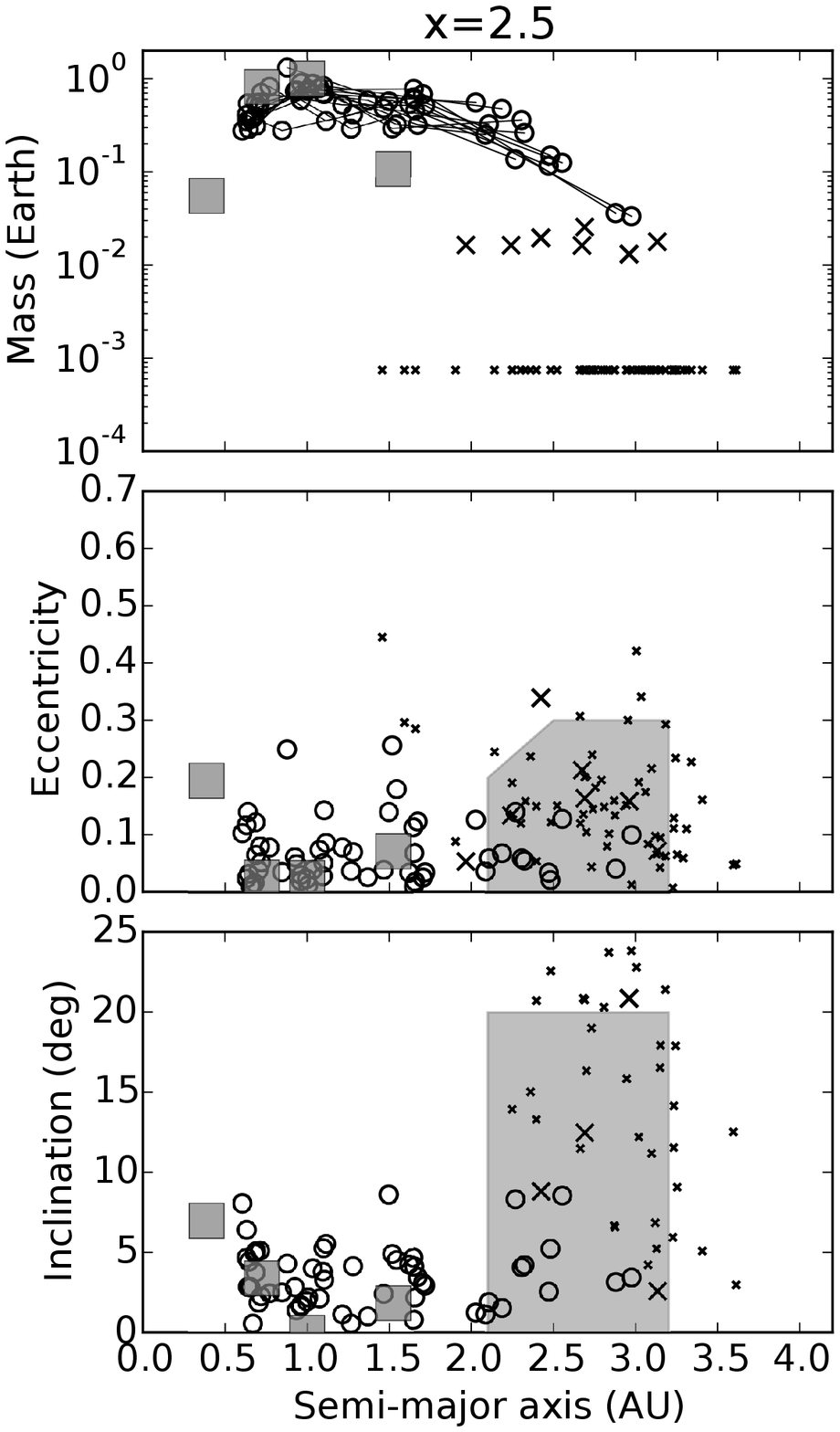} 
\includegraphics[width=0.45\textwidth]{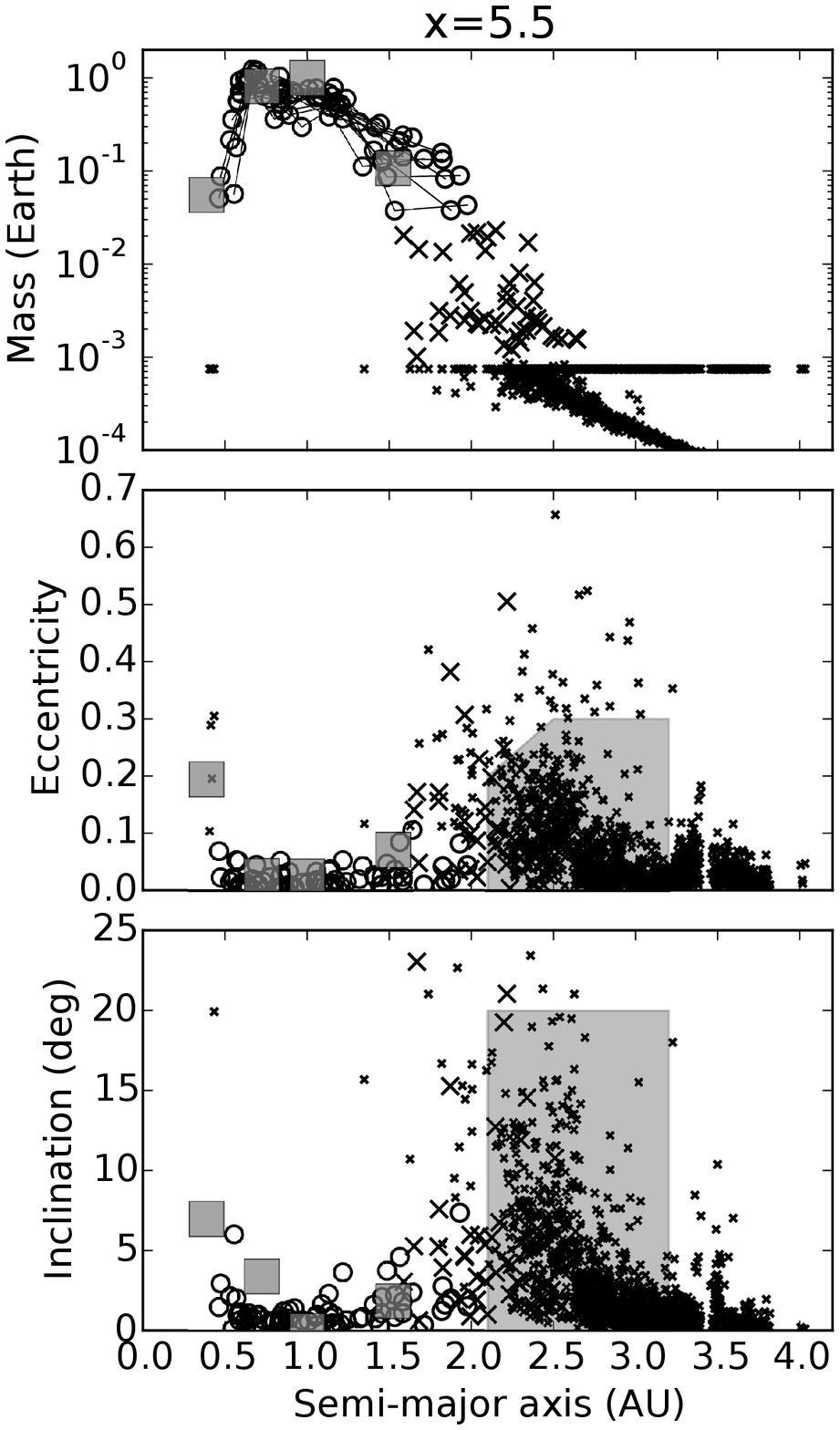} 
    \caption[]{The connection between asteroid belt excitation and terrestrial planet formation in the classical model. The three panels on the left come from a suite of simulations containing rocky planetary embryos and planetesimals within a disk following a surface density profile $\Sigma \propto r^{-x}$, where $x=2.5$, and the ones on the right are from analogous simulations in a disk with a much steeper radial surface density gradient ($x=5.5$). Note that the minimum-mass solar nebula model has $x= 1.5$~\citep{weidenschilling77,hayashi81} and observations of the outer parts of protoplanetary disks find $x=0.5-1$~\citep{andrews09,williams11}.  {\bf Top:} Mass vs orbital separation for the surviving planets after 700 Myr of evolution. {\bf Middle:} Eccentricity vs. orbital radius of all surviving bodies. {\bf Bottom:} Inclination vs. orbital radius for surviving bodies. Circles represent surviving planetary embryos or planets and dots are planetesimals. The main asteroid belt is shaded.The actual terrestrial planets are shown with solid triangles. Adapted from \cite{izidoro15c}.  }
         \label{fig:selfstirring}
         \end{center}
\end{figure}

Early studies showed that the primordial asteroid belt could have been excited to its current level in eccentricity and inclination by a population of Moon- to Mars-mass planetary embryos~\citep{wetherill92,chambers01b,petit01,obrien07}. There are constraints on this model.  First, embryos could not have survived in the belt for too long because they would open gaps in the distribution of asteroids that are not seen~\citep[see Fig.~1 in ][]{raymond09c}. These gaps could have been smeared out to some degree during the giant planet instability but only for certain evolutionary scenarios~\citep{brasil16}. Simulations of terrestrial planet formation in the context of the classical model often strand embryos in the belt~\citep[e.g.][]{raymond06b,raymond09c,fischer14b}. Second, the total amount of mass contained in the asteroid belt must have remained consistent with the mass distribution of the terrestrial planets (see below).

Figure~\ref{fig:selfstirring} shows how, in the self-stirring model, the level of excitation of the asteroids is linked with the mass of Mars~\citep{izidoro15c}. This is a consequence of the mass distribution of the belt. In simulations in a relatively shallow disk (i.e., with large asteroidal mass), massive embryos in the belt stirred up the eccentricities and inclinations of the asteroids, although embryos were often stranded within the belt. While this is inconsistent with the present-day Solar System, the more severe problem for shallow-disk simulations comes from the terrestrial planets.  These simulations suffer from the `small Mars problem'~\citep{wetherill91,raymond09c}: they systematically form Mars analogs that are far more massive than the real one. There is simply too much mass in Mars' feeding zone.

In simulations in a very steep disk (right-hand panels of Fig.~\ref{fig:selfstirring}) much less mass starts off in the asteroid belt and no embryos more massive than $10^{-3} \mearth$ form past roughly 2 au~\citep[see][for details]{izidoro15c}. Given the mass deficit in Mars' feeding zone these simulations form terrestrial planets that match the real ones, with Earth/Mars mass ratios comparable to the real one.  However, the mass deficit in the asteroid belt suppresses self-stirring such that the asteroids' eccentricities and inclinations are too low, especially in the outer main belt. 

This argument illustrates the no-win nature of the self-stirring model (at least in the context of the classical model of terrestrial planet formation; we will discuss alternatives in Sect. 4).  Either the belt is self-excited but Mars is far too massive or the correct Earth/Mars mass ratio is matched but the asteroid belt is under-excited~\citep{izidoro15c}.

\subsection{Implantation of asteroids from the terrestrial planet region}

There is ample circumstantial evidence that the terrestrial planets grew from a population of $\sim$Mars-mass planetary embryos~\citep[see discussion in][]{morby12b}.  During this process -- which lasted at least until Earth's last giant impact at 50-100 Myr~\citep{kleine09,jacobson14,bottke15} -- planetesimals were frequently scattered in all directions by planetary embryos.  Most planetesimals ended up being accreted by a growing planet or ejected after a close encounter with Jupiter~\citep{raymond06b}, but a small fraction were implanted into the asteroid belt.

\cite{bottke06} proposed that the parent bodies of iron meteorites were implanted from the terrestrial planet-forming region. They found that roughly one in a thousand terrestrial planetesimals was scattered outward by successive encounters with embryos and ended up on a stable orbit within the belt~\citep[see also][]{haghighipour12,mastrobuonobattisti17}. While the mechanism of implantation was conceptually sound, these studies piggybacked on the classical model of terrestrial planet formation and therefore suffered from the same fatal flaw. Implantation relied on a chain of embryos that scatter planetesimals outward from roughly 1 au to the main belt. Those same chains of embryos formed Mars analogs that were far too massive~\citep[once again, the dreaded `small Mars' problem][]{wetherill91,raymond09c}.

\cite{hansen09} proposed that the initial conditions being used in classical model simulations were not the real ones.  He showed that the large Earth/Mars mass ratio was naturally matched if the terrestrial planets did not form from a broad disk but from a narrow annulus from 0.7-1 au.  Mars and Mercury were scattered out of the annulus and starved whereas Earth and Venus grew massive within the annulus~\citep[see also][]{kaib15,raymond17b}. More detailed simulations find that this evolution may require embryos to form very quickly within the annulus~\citep[or the annulus to have shaped by external forces as in the Grand Tack model; see][]{jacobson14b}; otherwise, a narrow annulus of planetesimals rapidly spreads out and systems that emerge often fail to match the real terrestrial planets using metrics such as the {\it radial mass concentration}~\citep{walsh16,deienno19}. 

Within the framework of a narrow annulus, \cite{raymond17b} showed that planetesimals are implanted into the asteroid belt from the terrestrial planet-forming region. While the efficiency of implantation was only $\sim 10^{-3}$ this was sufficient to implant all of the S-types, even accounting for later dynamical losses (e.g., during the giant planet instability). It is interesting to note that the rate of implantation of terrestrial planetesimals from a narrow annulus~\citep[as in][]{raymond17b} is only modestly smaller than from a broad disk~\citep[as in][]{bottke06}.  

\begin{figure}
\begin{center}
\includegraphics[width=0.7\textwidth]{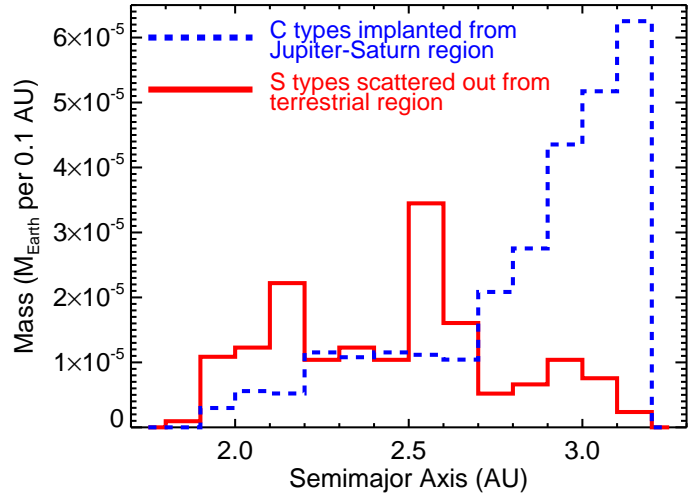} 
    \caption[]{Radial distribution of a 100\% implanted asteroid belt~\citep[adapted from][]{raymond17b}. The red curve indicates asteroids scattered into the main belt from the terrestrial planet-forming region, and the blue dashed curve are asteroids scattered inward during Jupiter and Saturn's growth (by the mechanism illustrated in Fig.~\ref{fig:injection}). }
         \label{fig:empty_ab}
         \end{center}
\end{figure}

Implantation of asteroids from a terrestrial annulus is slightly different than from a broad disk. Figure~\ref{fig:empty_ab} shows the distribution of implanted terrestrial planetesimals from~\cite{raymond17b}. Planetesimals are continually scattered by growing embryos onto eccentric orbits that may cross the asteroid belt. These orbits have semimajor axes within the asteroid region but their eccentricities are too high. Two mechanisms can drop their eccentricities (under gas-free conditions). First, a planetesimal can be scattered into a strong mean motion resonance with Jupiter, which causes an angular momentum exchange that can drop the planetesimal's eccentricity. Planetesimals in or near the 3:1 resonance with Jupiter can be seen in Fig.~\ref{fig:empty_ab} as the peak at $\sim 2.55$~au. Resonant planetesimals with high eccentricities are not stable for long timescales unless another event can drop their eccentricities further, for example the Kozai resonance within mean motion resonances, which offers a robust path to low eccentricity. A planetesimal trapped on a resonant orbit with Jupiter when the giant planet instability took place may have ended up on a long term stable orbit if it was removed from resonance as Jupiter's orbit shifted (see Sect. 3.4 below). 

The second mechanism for trapping scattered planetesimals relies on embryo scattering. During accretion embryos scatter each other onto asteroid belt-crossing orbits, sometimes crossing Jupiter's orbit (often leading to ejection from the Solar System).  When a scattered embryo encounters a scattered planetesimal with a random phase, the planetesimal's eccentricity can either increase or decrease. If the eccentricity increases the planetesimal crosses Jupiter's orbit and is quickly ejected. But if the eccentricity decreases the planetesimal may be trapped on a stable, non-resonant orbit within the main belt. This mechanism was the most effective in implanting terrestrial planetesimals into the belt in the simulations from Fig.~\ref{fig:empty_ab}. Scattered embryos do not survive in the belt but end up either colliding with a growing terrestrial planet or being ejected. 

A primordial empty asteroid belt would have been populated with scattered planetesimals that roughly match the present-day belt's radial distribution. Fig.~\ref{fig:empty_ab} shows the radial distribution of planetesimals implanted as a byproduct of the giant planets' growth from a suite of simulations like the one from Fig.~\ref{fig:injection}. This matches the compositional distribution of the belt if we associate terrestrial planet-forming planetesimals with S-types and giant planet-region planetesimals with C-types.    

It is not clear that planetesimals from the terrestrial planet-forming region should be associated with S-types.  Ordinary chondrite meteorites~\citep[sourced from S-type asteroids;][]{bus02} are compositionally different than Earth~\citep[e.g][]{warren11}. While chemical models for Earth's growth allow for a contribution from ordinary chondrites they generally find that Earth is mostly made from Enstatite chondrite-like material~\citep[e.g.][]{dauphas17}. Given that the dynamical mechanism of planetesimal implantation is robust, it is unclear whether S-types are indeed representative of the terrestrial planet-forming region but the terrestrial planetesimals' compositions varied in time and/or with radial distance~\citep[e.g.][]{schiller18}, or whether implanted planetesimals simply only represent a fraction of inner belt asteroids (perhaps Enstatites).

\subsection{Chaotic excitation by Jupiter and Saturn}

\cite{izidoro16} showed that chaos in the orbits of Jupiter and Saturn could have excited the asteroids' orbits.  Emerging from the gaseous disk, Jupiter and Saturn's orbits sometimes exhibit chaotic evolution.  This is more likely if Jupiter and Saturn were in 2:1 resonance rather than in 3:2, although chaotic configurations exist throughout the relevant parameter space.  

\begin{figure}
\begin{center}
\includegraphics[width=0.49\textwidth]{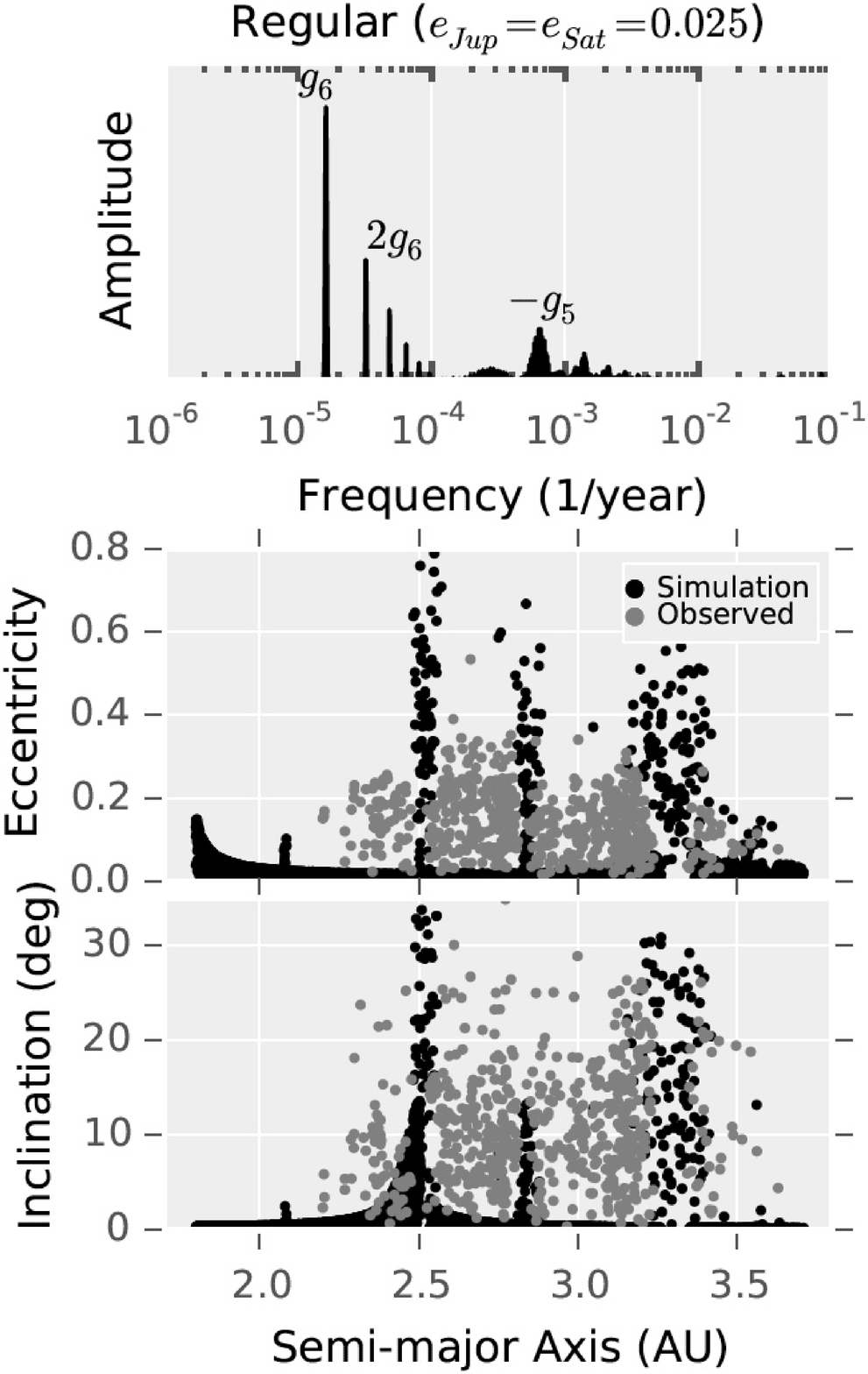} 
\includegraphics[width=0.49\textwidth]{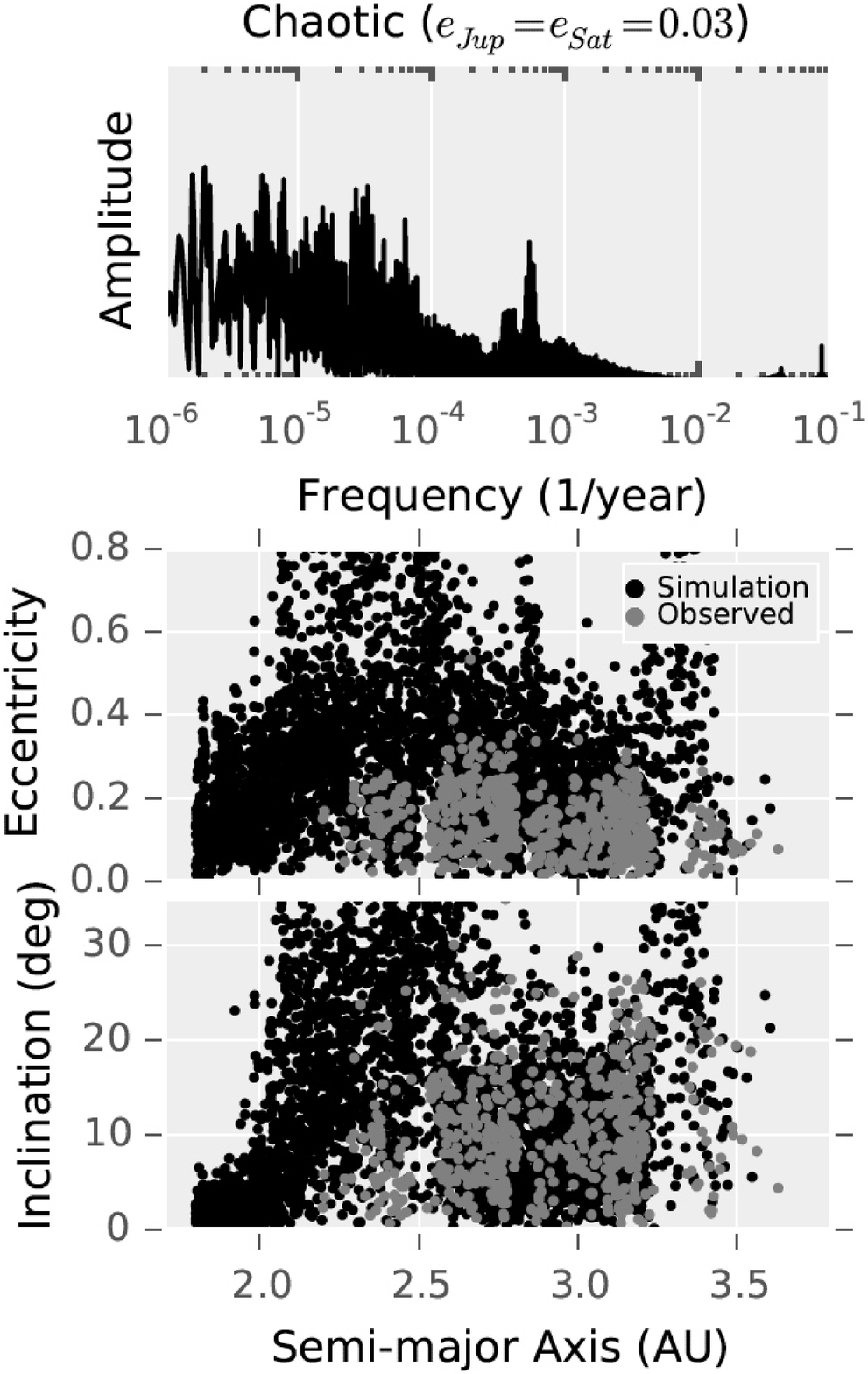} 
    \caption[]{Chaotic excitation of the asteroid belt~\citep[adapted from][]{izidoro16}.  {\bf Left:} An example simulation in which Jupiter and Saturn's evolution is regular, with the planets locked in 2:1 resonance as an outcome of their migration within the gaseous disk~\citep{pierens14}. {\bf Right:} An example simulation in which Jupiter and Saturn's eccentricity is only slightly higher (0.03 rather than 0.025) than in the panels on the left but their evolution is chaotic. The top panels show the frequency spectrum of each system and the bottom panels show the asteroid belt's inclinations and eccentricities, as compared with the present-day main belt (in gray).  Chaos in the giant planets' orbits causes secular resonances to have a broad frequency spectrum and act to excite asteroids across the entire belt rather than only at specific orbital radii.}
         \label{fig:chaos}
         \end{center}
\end{figure}

Figure~\ref{fig:chaos} shows how the asteroids are excited when the giant planets exhibit regular or chaotic motion. In the left-hand simulation, the peaked frequency spectrum indicates that the system is dominated by a small number of fundamental modes and that Jupiter and Saturn's orbits are evolving regularly. Secular resonances within the belt -- corresponding to specific locations within at which asteroids' precession rates match these frequencies -- remain fixed.  Asteroids' eccentricities and inclinations are only excited in those locations. This does not reproduce the observed distribution (shown as the grey points in the bottom two panels), with asteroids excited across the width of the belt (see also Fig.~\ref{fig:mainbelt}). 

In contrast, when Jupiter and Saturn evolve chaotically they excite the entire width of the asteroid belt~\citep{izidoro16}. In the right-hand simulation of Fig.~\ref{fig:chaos}, Jupiter and Saturn's initial orbits evolved chaotically, as shown by the broad (not peaked) precession frequency spectrum. The planets' precession frequencies underwent chaotic jumps, causing the locations of secular resonances to jump to different orbital radii. This acts to excite the full width of the belt. In the example from Fig.~\ref{fig:chaos} the belt may even have been over-excited, as there is a large population of asteroids with inclinations above $20^\circ$ that cannot easily be removed~\citep{roig15,deienno16,deienno18}.

For this chaos-driven mechanism to be responsible for exciting the belt, two conditions must apply.  First, the belt's mass must initially have been low.  If the primordial belt was massive then perturbations between objects would overwhelm the secular resonant perturbations. Of course, if the primordial belt contained a lot of mass then it would simply have self-excited (Fig.~\ref{fig:selfstirring}). Second, the giant planets must have spent enough time in a chaotic configuration to excite the whole belt. It is plausible to imagine that the giant planets emerged from the gaseous disk in a chaotic configuration (with a low-mass asteroid belt) and then excited the asteroids' orbits during an interval of at least a few Myr before the onset of the giant planet instability~\citep{izidoro16}.

\subsection{Effect of the giant planet instability: dynamical excitation, radial mixing and implantation}

It is now well-accepted that the giant planets underwent a dynamical instability after their formation~\citep[see review by][]{nesvorny18}. The giant planet instability offers an elegant explanation for many small body Solar System populations and has important consequences for the asteroid belt.  In the upcoming subsections we first discuss the instability itself and then the implications of the instability for the asteroid belt. We discuss the implantation of outer Solar System planetesimals, the depletion and excitation of the belt and the smearing out of asteroid families.  Finally, we discuss how the final phase of spreading out of Jupiter and Saturn's orbits appears to have left its imprint in the inclination distribution of the inner belt asteroids.

\subsubsection{The giant planet instability}

The idea that the giant planets' orbits have changed since their formation was inferred by \cite{fernandez84} by considering the back-reaction of planetesimals scattered into the inner Solar System. This is called planetesimal-driven migration. Neptune, Uranus or Saturn are not massive enough to scatter planetesimals to hyperbolic orbits~\citep[e.g.][]{duncan87}. Planetesimals instead are scattered inward by Neptune, Uranus and Saturn, and then Jupiter scatters them out to interstellar space. Orbital energy conservation dictates that Neptune, Uranus and Saturn migrate outward whereas Jupiter migrates in. \cite{malhotra93,malhotra95} showed that Pluto's resonant orbit with Neptune could be taken as evidence for Neptune's outward migration.  

Planetesimal-driven migration likely took place in the early Solar System. It was originally suggested~\citep{levison01,gomes05} that migration was delayed in order to explain the LHB (Late Heavy Bombardment), an apparent spike in the bombardment in the inner Solar System~\citep{tera74,gomes05,morby12d,marchi12,bottke12}. Constraints on the collisional grinding of the outer disk indicate, however, that there probably was no delay and migration started no later than 100 Myr after the gas disk dispersal~\citep{nesvorny18b}. Lunar craters were probably produced by impactors from the terrestrial planet region~\citep{morby18}.

The timescale of planetesimal-driven migration can be constrained from the structure of the asteroid and Kuiper belts. If Jupiter's planetesimal-driven migration were slow, secular resonances would produce a large population of asteroids on highly inclined orbits~\citep[$i>15^\circ$][]{morby10,walsh11b}, and would destabilize the orbits of the terrestrial planets~\citep{brasser09,agnor12}. The absence of any large high-$i$ population in the present asteroid belt indicates that Jupiter's (and Saturn's) migration was fast (e-folding time $\tau < 1$ Myr). In contrast, the distribution of orbital inclinations in the Kuiper belt is broad, reaching above 30$^\circ$. This implies that Neptune's migration was slow~\citep[with e-folding time $\tau > 5$ Myr, ][]{nesvorny15}. This is a problem because the migration timescales of Jupiter and Neptune cannot be so different, as they are both linked to the outer disk mass~\citep[more massive disks produce faster migration; ][]{gomes04}.

The tension between asteroid and Kuiper belt constraints can be resolved if a dynamical instability took place. This idea is sometimes called the {\em Nice model} because it was first developed in Nice, France~\citep{tsiganis05}. Its most recent incarnation is sometimes called the {\em Jumping Jupiter} model.

The current paradigm for the giant planet instability is as follows~\citep[see][for a review]{nesvorny18}. The giant planets likely emerged from the gaseous disk in a multi-resonant chain, a generic outcome of planet-disk interactions~\citep[e.g.][]{morby07}. Jupiter and Saturn were either in 3:2 or 2:1 resonance~\citep{pierens14}. An outer disk of planetesimals extended out past 30 au and contained a total of $\sim 15-20\mearth$. The instability may have been triggered by a resonance between a pair of migrating planets~\citep[see][]{levison11,deienno17,nesvorny18,quarles19}, although it is possible that the giant planet system was unstable on its own~\citep[meaning that interactions with the planetesimal disk may not have been the trigger of instability -- ][although the initial conditions, and thus the formation models, are critical in determining self-instability timescales]{ribeiro20}. While the exact timing of the instability is unknown, it almost certainly took place within 100 Myr of CAIs~\citep{nesvorny18b,mojzsis19}. During the instability at least one ice giant was scattered by Jupiter and likely ejected~\citep[meaning the Solar System likely formed 1-2 extra ice giants][]{nesvorny11,nesvorny12,batygin12}. The semimajor axis of Jupiter nearly instantaneously changed during the scattering event and the secular resonances jumped from $>$3.5 au to $\sim2$ au (this is the `jumping Jupiter' variant of the Nice model). This resolves the problem with the high-$i$ population because the $\nu_{16}$ resonance never spent too much time between 2 and 3.5 au~\citep{morby10}.  The instability can also explain why Jupiter has a significant orbital eccentricity (proper $e=0.044$). The outer planetesimal disk was destabilized, and the Oort cloud and Kuiper belt were populated~\citep{levison08,brasser13b,nesvorny15}.  As the last dregs of the planetesimal disk were cleared giant planets reached their present-day orbital configuration. {\em In summary, the planetesimal-driven migration could have been slow as required from the Kuiper belt constraints, but the jumping Jupiter instability is needed for the asteroid belt~\citep[see][]{nesvorny15,deienno17}.}

\subsubsection{Implantation of D-types from the trans-Neptunian disk}

During the gaseous disk phase, the giant planets' growth and migration acted to implant bodies from $\sim$5-20 au into the asteroid belt (see discussion in Sect. 3).  A second stage of implantation followed after the gas disk dispersed, as a consequence of the giant planet instability. 

\begin{figure}
\begin{center}
\includegraphics[width=0.8\textwidth]{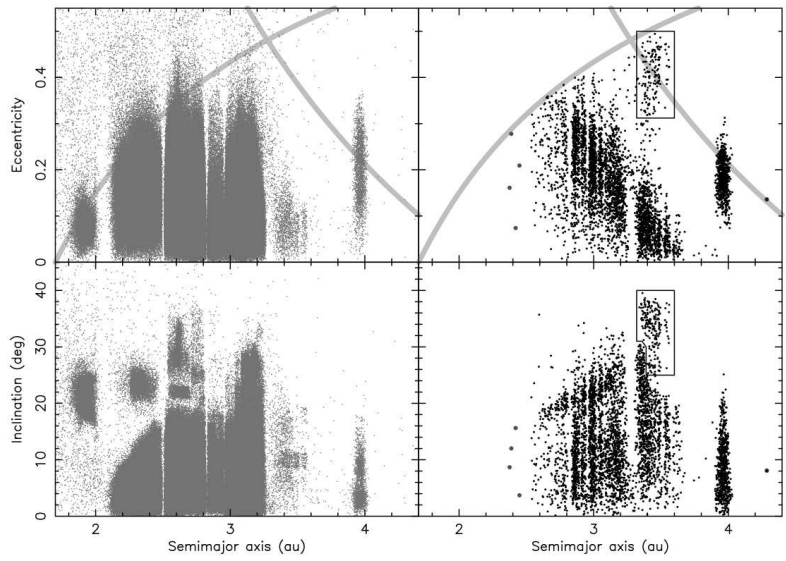} 
    \caption[]{Population of asteroids captured from the outer planetesimal disk (right) in the giant planet instability simulations of~\cite{vokr16}, compared with the actual distribution of all numbered asteroids (left). The four innermost (larger) particles were implanted interior to the 3:1 resonance with Jupiter. The rectangle/polygon in the right panels indicates the population of particles captured in Kozai states~\citep[for details, see][]{vokr16}.}
         \label{fig:dtypes}
         \end{center}
\end{figure}

By examining simulations of the original giant planet instability model, \cite{levison09} noticed that a fraction of outer disk planetesimals ended on stable orbits in the asteroid belt. They hypothesized that planetesimals were captured and subsequently released from migrating resonances with Jupiter, leaving planetesimals on low-$e$ orbits in the main belt. The results suggested that the implanted population could be huge, some $\sim1$-2 orders in magnitude larger than the current population of main-belt asteroids. \cite{vokr16} used a large statistical sample (effectively $\sim10^9$ bodies) to revisit this issue. Using the best migration/instability models from \cite{nesvorny12}, they confirmed the general process of implantation proposed by \cite{levison09} but implanted a smaller fraction of the outer disk planetesimals.  \cite{vokr16} assumed the outer disk to start with $\sim 5 \times 10^7$ bodies with $D>100$ km and a Trojan-like size distribution for $D<100$ km (differential power law index $q\simeq2$). Figure~\ref{fig:dtypes} shows the orbital distribution of asteroids captured in the simulations from \cite{vokr16}. 

The outer disk planetesimals are thought to be represented by P/D types in asteroid taxonomy. This is because simulations of the giant planet instability also explain Jupiter's Trojan asteroids -- which are predominantly D/P types -- as captured outer disk planetesimals~\citep{morby05,nesvorny13}. By comparing the implanted population with the main belt P/D-types, \cite{vokr16} found that: (1) the number of {\it large} P/D bodies implanted in the model matches observations, but (2) the population of implanted P/D bodies with $D \sim 10$ km is roughly 10 times larger than observed. \cite{vokr16} argued that collisional grinding, thermal destruction of bodies before their implantation and/or removal of small bodies by the Yarkovsky effect would resolve this issue. This remains to be demonstrated. The implantation hypothesis described here raises an interesting possibility that some D-type Near-Earth Asteroids, potential targets of sample-return space missions, are compositionally related to Kuiper belt objects.

\subsubsection{Dynamical excitation and depletion}

The giant planet instability's effect on the asteroid belt depends on the belt's initial distribution and the exact evolution of the planets. A common assumption is that the primordial asteroid belt was massive and dynamically cold, in contrast with the present-day belt's low mass and excited orbits (see Sect. 1.1). However, it is entirely possible that the primordial belt was already depleted and that a weaker instability could still explain the present-day belt.

Recent studies have simulated the dynamics of the inner Solar System during and after the giant planet instability. \cite{roig15} modeled the asteroid belt's dynamical evolution during the instability including just the giant planets, and \cite{nesvorny17b} extended this analysis to also include the terrestrial planets, assuming them to be fully-formed at the time of instability.  Both treated the asteroids as massless test particles initially spread across a broad range of parameters space ($a$, $e$ and $i$) without specifying the initial mass. Instead, the asteroid orbits were propagated over 4.5 Gyr and matched against the present asteroid belt. The results were used to estimate of the overall depletion and the initial mass. A small number of orbital histories for the giant planets were carefully selected from a vast number of simulations \citep{nesvorny12} to match various Solar System constraints (e.g., the outer planet orbits, Kuiper belt, planetary satellites, Jupiter Trojans). The planets' migration histories were strictly controlled to only consider orbital evolutions compatible with constraints.  Both studies accounted for the effects of an additional ice giant whose orbit may have briefly ($\sim$10,000 yr) overlapped with the asteroid belt during the instability. The selected migration/instability models were compatible with the dynamical structure of the terrestrial planets, including the excited orbit of Mercury~\citep{roig16}. Thus, whereas the terrestrial planet system is in general fragile and susceptible to excessive dynamical excitation during planetary migration and instability~\citep{agnor12,kaib16}, the cases considered in \cite{roig15} and \cite{nesvorny17b} did not have this problem.

\begin{figure}
\begin{center}
\includegraphics[width=0.7\textwidth]{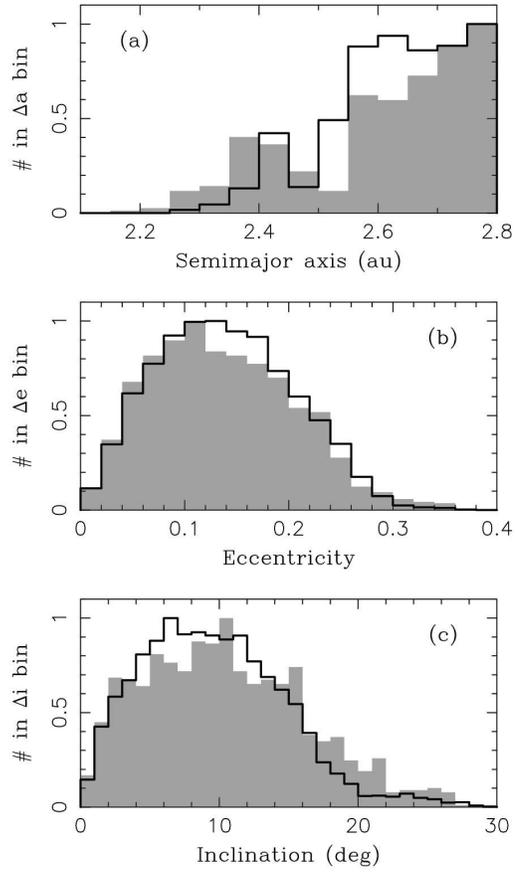} 
    \caption[]{A comparison of the orbital distribution of $D>30$~km main belt asteroids (histograms) with the distribution from simulations by \cite{nesvorny17b} of asteroid belt sculpting during the giant planet instability and for an additional 4.5 Gyr (dark lines). }
         \label{fig:nesvorny17}
         \end{center}
\end{figure}

Figure~\ref{fig:nesvorny17} shows the final distribution of simulated asteroids, which provide a good match to the present-day belt. \cite{roig15} and \cite{nesvorny17b} did not find a difference in the final orbital distribution of asteroids as a function of the timing of the instability. In their setup there was not much difference between the effects of dynamical depletion over $\sim$4.5 Gyr (for the early instability) or over $\sim$4 Gyr (for the late instability). The early instability led to a slightly smaller number of asteroids surviving at the present time, because the dynamical erosion with Jupiter on its current orbit has more time available. The original (i.e., at the time of the solar nebula dispersal) inclination distribution of main belt asteroids was inferred to have been relatively narrow with a great majority of bodies starting with $i<20^\circ$~\citep{roig15}. This is because objects starting with $i>20^\circ$ would have survived, but a large population of main belt asteroids with $i>20^\circ$ does not exist today~\citep{morby10}. The nominal Grand Tack model produces an inclination distribution that is slightly too wide to satisfy this constraint \citep{deienno16} but it is possible that a minor modification of the original Grand Tack model would yield a more satisfactory result.

The original eccentricity distribution is less well constrained because planets act, though overlapping orbital resonances, to erode the population of high-$e$ objects. For example, \cite{deienno16} showed that the broad eccentricity distribution produced by the Grand Tack model would erode toward a final state similar to the present asteroid belt. Indeed, \cite{roig15} favored cases where the belt was at least slightly excited in eccentricities before the planetary migration/instability happened. \cite{minton11} considered the eccentricity evolution of asteroids during planetary migration (without instability) and found that either the belt was initially dynamically cold and Jupiter/Saturn migrated fast, or the belt was dynamically hot and Jupiter/Saturn migrated slow. It is not clear what the meaning of this result is in the jumping Jupiter paradigm. Finally, \cite{deienno18} showed that a subset of jumping Jupiter models with more chaotic evolution of Jupiter and Saturn can strongly excite the main belt. The mechanism demonstrated in \cite{deienno18} is similar to the chaotic excitation mechanism of \cite{izidoro16} discussed in Sect. 3.3; it relies on secular excitation of asteroids by Jupiter and Saturn while their orbits are themselves in flux during the instability. 

The degree of depletion of the belt remains an area of contention. \cite{nesvorny17b} found that $\sim$20\% of the original population of asteroids (starting at $\sim$2-3.5 au) would survive in the main belt today. This estimate applies to the model of planetary migration/instability that best satisfies Solar System constraints, includes the dynamical erosion of the asteroid belt over 4.5 Gyr, and treats the asteroids as test particles. Models without planetary migration/instability indicate larger surviving fractions. For example, \cite{minton11} estimated that $\sim$50\% of main belt asteroids would have survived since $\sim$ 1 Myr after the establishment of the current Solar System architecture. This is consistent with the results obtained in \cite{nesvorny17b} in the models {\it without} planetary migration/instability. Most removed asteroids are therefore expected to be removed early.

A higher fraction ($\sim 90\%$) of asteroids is removed during the instability starting from the orbital distributions obtained from the Grand Tack model (see Sect. 2.2.2)~\citep{deienno16}. This is due to a large unstable population of asteroids on high-$e$ orbits. Still, this depletion factor is not large enough to explain the main belt depletion, if the main belt started massive. The combined depletion from the Grand Tack (99\%) and subsequent migration/instability would work, within an order of magnitude, to reduce an initially massive main belt to the present population~\citep{deienno16}. The asteroid belt depletion is not a problem if the asteroid belt started with a low mass (see discussion in Sect. 4.1). It is interesting that the inner belt ($a<2.5$ au) contains $\sim$10 times less mass that the rest of the main belt despite containing roughly one third of the orbital real estate. This can be explained if the secular resonances such as $\nu_6$ and $\nu_{16}$ spent some time at 2-2.5 au and strongly depleted the inner belt, which indeed takes place in the jumping-Jupiter model~\citep{nesvorny17b}.

Simulations in which the asteroids carry mass sometimes find larger depletion factors than those that treat asteroids as massless test particles. \cite{clement19a} showed that depletion of up to 3-4 orders of magnitude is possible when taking into account the pre-excitation of the belt by self-stirring. A caveat is that the pre-excitation in \cite{clement19a} was probably an upper limit on the plausible range as it was driven by gravitational self-stirring of very massive planetary embryos. In addition, the strongest depletion factors came from simulations that did not provide good matches to the Solar System, for example with the giant planets on orbits that are much more excited or spread out than the actual ones. Yet depletion factors of two orders of magnitude ($\sim$99\% of asteroids removed) were a common outcome among systems that matched constraints. This also fits into models of terrestrial planet formation, as \cite{clement18,clement19b} showed that if the giant planet instability took place early enough it could also explain the terrestrial planets' mass distribution (see Sect. 4). 

What are we to make of the differences between models of the inferred asteroid depletion during the giant planet instability?  There are two fundamental differences between the studies. First, \cite{roig15}, \cite{nesvorny17b} and \cite{deienno18} imposed a specific orbital history for the giant planets that was drawn from simulations of the instability itself that best matched constraints~\citep{nesvorny12}. They then studied the effects of the giant planets' dynamics on the asteroids.  In contrast, \cite{clement19a} did not prescribe the giant planets' evolution but only triggered their instability, then simulated its effects. Second, while \cite{roig15} and \cite{nesvorny17b} treated the asteroids as massless test particles, \cite{clement19a} treated them as massive objects that interacted gravitationally with each other (assuming them to be equal-mass and quite massive) and with the giant planets. 

Dynamical instabilities are inherently chaotic. Infinitesimal changes in the configuration of individual close encounters between planets may completely change the outcome. There is no single evolution that can produce the Solar System but rather a wide range of possible pathways. The studies of \cite{roig15}, \cite{nesvorny17b} and \cite{deienno18} only included giant planet evolutionary pathways that are consistent with constraints such as avoiding the over-excitation of the asteroids and insuring the survival of the (assumed to be already-formed) terrestrial planets. However, they only included a tiny subset of possible giant planet evolutionary pathways. In contrast, \cite{clement19a} sampled the effects of a much broader range of instabilities, but that included many that were not consistent with the present-day Solar System. The differences in outcome between the studies show how sensitive the asteroid belt's dynamics is to the exact evolution of the giant planets during the instability. Each of these approaches is valid: \cite{roig15}, \cite{nesvorny17b} and \cite{deienno18}'s approach goes {\em deep} to study the detailed consequences of a small number of plausible instabilities whereas \cite{clement19a} goes {\em wide} and captures the outcome of more general instabilities, essentially putting the Solar System's instability in a larger context.

The two sets of studies' outcomes may in principle be reconciled by simply taking into account the asteroid belt's mass distribution. Imagine a scenario in which the early asteroid belt included ten large planetary embryos that contained 90\% of the belt's mass. In this idealized case, and assuming that all ten embryos were lost during the instability, removal of 90\% of objects within the belt would produce a mass depletion of two orders of magnitude. 

The mixing of asteroids in semimajor axis during the planetary migration/instability was insufficient (characteristic change $\Delta a < 0.1$ au) to explain the compositional mixing of the asteroid belt~\citep{roig15,clement19a}. Compositional mixing therefore requires a stronger effect such as the ones discussed in Sect. 2. Collisional families that should have formed during an intense period of collisional activity after the gas nebula removal would have been dispersed~\citep{brasil16,brasil17}. The dispersal of asteroid families in $e$ and $i$ is especially significant suggesting that most ancient families cannot be identified by the usual clustering techniques. The V-shape method seems to be more promising~\citep{delbo17}. \cite{minton09} identified gaps just outside the 5:2, 7:3 and 2:1 resonances and explained them by invoking a dynamical model with fast migration of Jupiter (one e-fold $\simeq$0.5 Myr; no instability). The gaps are created in their model when orbital resonances with Jupiter move inward and remove objects in their path. It may be difficult to obtain the same result with the jumping-Jupiter model where resonances jump to their final locations.

\subsubsection{Effect of Jupiter and Saturn's final migration phase}

A common shortcoming of models for inner Solar System formation is that they produce too many high-inclination asteroids in the inner main belt~\citep{obrien07,morby10,walsh11b,roig15,deienno16,deienno18,clement18,clement19a}.  \cite{clement20} showed that this deficit may be a signature of the final phase of migration of Jupiter and Saturn. This phase took place after the giant planets' instability, as the last remnants of the outer planetesimal disk were scattered away, causing the giant planets' orbits to slowly spread apart.  This phase left an imprint on the asteroid belt via the $\nu_6$ secular resonance, where asteroids' orbits precess at the same rate as Saturn. The $\nu_6$ is a function of asteroid semimajor axis, eccentricity and inclination~\citep{morby91}.  In the inner main belt, the $\nu_6$ is located at $\sim 10-15\deg$ in inclination. Asteroids that fall in the $\nu_6$ are removed by being driven to high enough eccentricity to collide with the Sun~\citep{gladman97}. 

\begin{figure}
\begin{center}
\includegraphics[width=0.49\textwidth]{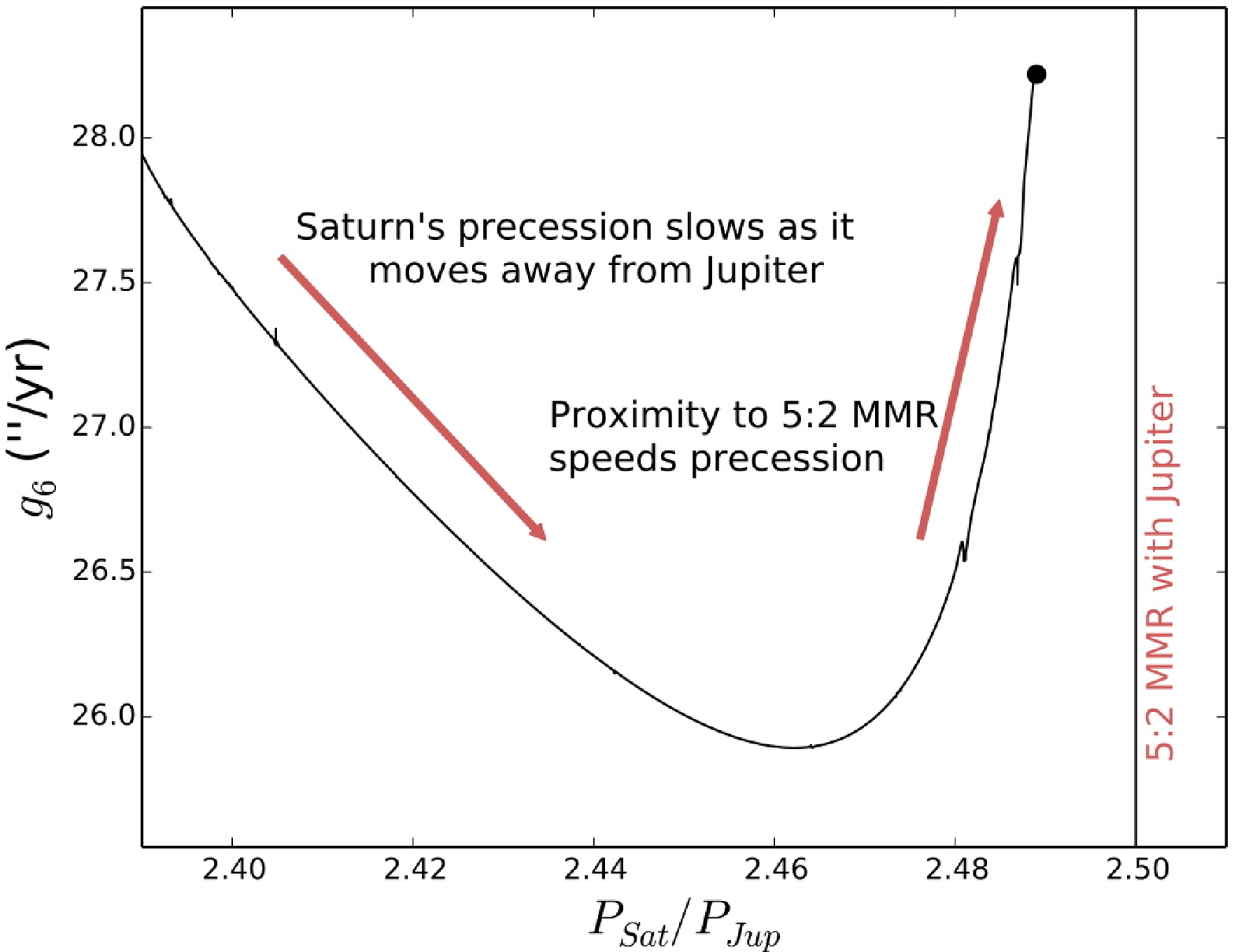} 
\includegraphics[width=0.49\textwidth]{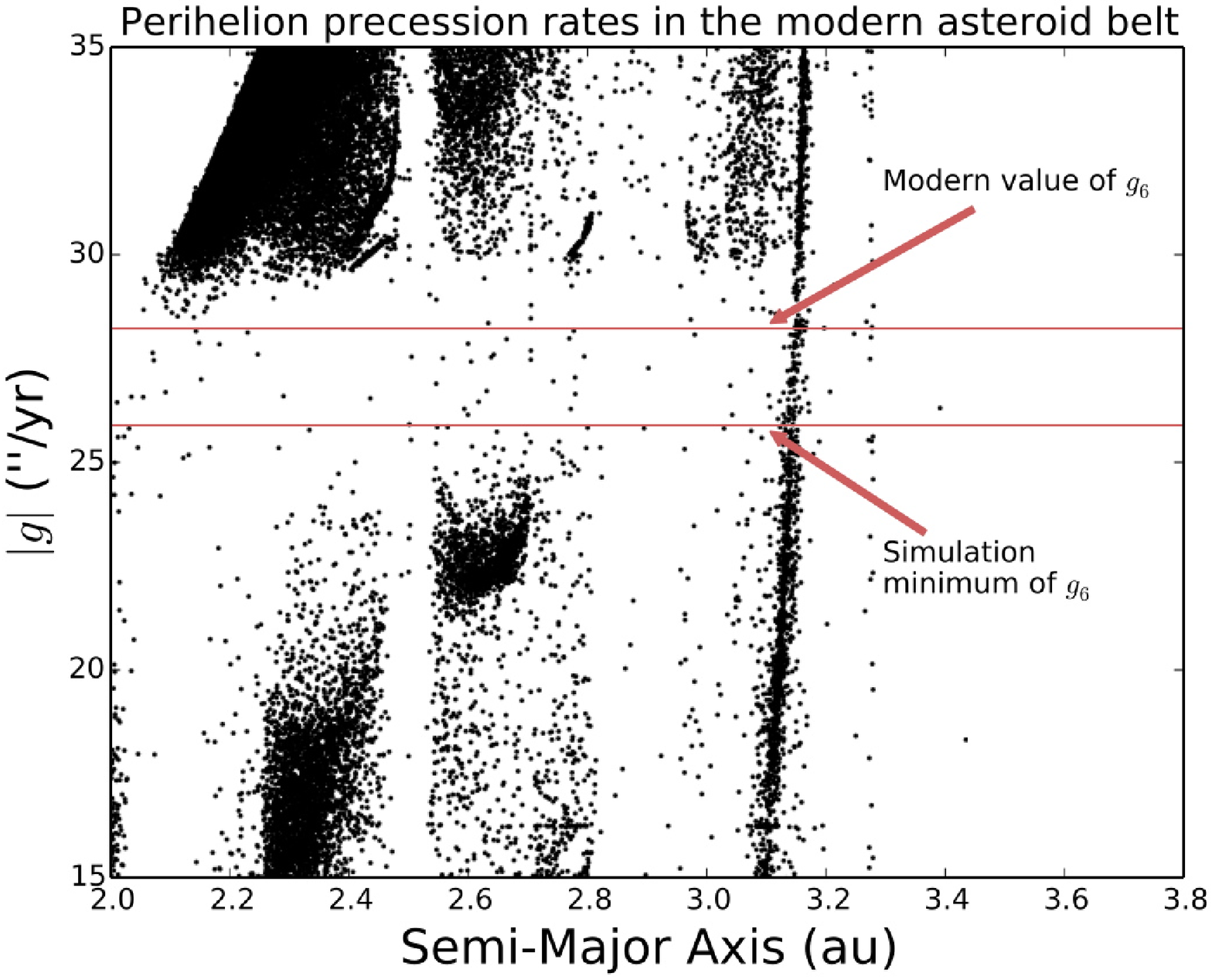} 
    \caption[]{How Jupiter and Saturn's final phase of migration shaped the asteroid belt~\citep[adapted from][]{clement20}.  {\bf Left:} Saturn's orbital precession rate as a function of its separation from Jupiter.  During the final phase of planetesimal scattering, Saturn moved away from Jupiter (from left to right in the figure).  The black dot represents the present-day configuration. {\bf Right:} Precession rates of known asteroids as a function of semimajor axis~\citep[from][]{knezevic03}.  The horizontal lines denote the minimum and maximum precession frequencies from the left panel.  The fact that almost no asteroids have precession rates within that range is evidence that they were dynamically removed by the $\nu_6$ secular resonance with Saturn during the giant planets' migration. The grouping of asteroids near 3.1 au that span the gap represents the Euphrosyne family~\citep{novakovic11} that is dynamically spreading due to the Yarkovsky effect~\citep[e.g.][]{bottke01}. }
         \label{fig:clement2}
         \end{center}
\end{figure}

Saturn's current precession rate is 28.22 arc seconds per year. Figure~\ref{fig:clement2} (left panel) shows how Saturn's precession rate would have evolved during its last phase of migration, as it distanced itself from Jupiter.  As Jupiter and Saturn's orbits spread apart, Saturn's precession rate naturally slowed.  However, after reaching a minimum of 26 arcsec yr$^{-1}$ its precession rate sped back up as it approached the 5:2 mean motion resonance with Jupiter~\citep{milani90,morby91}.  The right-hand panel of Fig.~\ref{fig:clement2} shows the distribution of asteroid precession rates. No asteroids can survive long at the same precession rate as Saturn because that would put them in the unstable $\nu_6$ secular resonance. If Saturn had always maintained its current orbit, then there should be a narrow gap in the distribution with no asteroids having precession rates of 28.2 arcsec yr$^{-1}$. \cite{clement20} interpreted the large empty swath at precession rates between 26 and 28 arcsec yr$^{-1}$ as evidence that Saturn's precession rate had swept between those values during its final phase of migration. 

The asteroids that were destabilized during the final phase of migration -- with precession rates within that empty range -- include those on high-inclination orbits in the inner parts of the main belt. Taking the final phase of migration into account reconciles models of inner Solar System dynamics (that did not force the giant planets to end up on their exact current orbits) with the orbital structure of the present-day belt~\citep{clement20}.

\section{Discussion} 

\begin{figure}
\begin{center}
\includegraphics[width=0.95\textwidth]{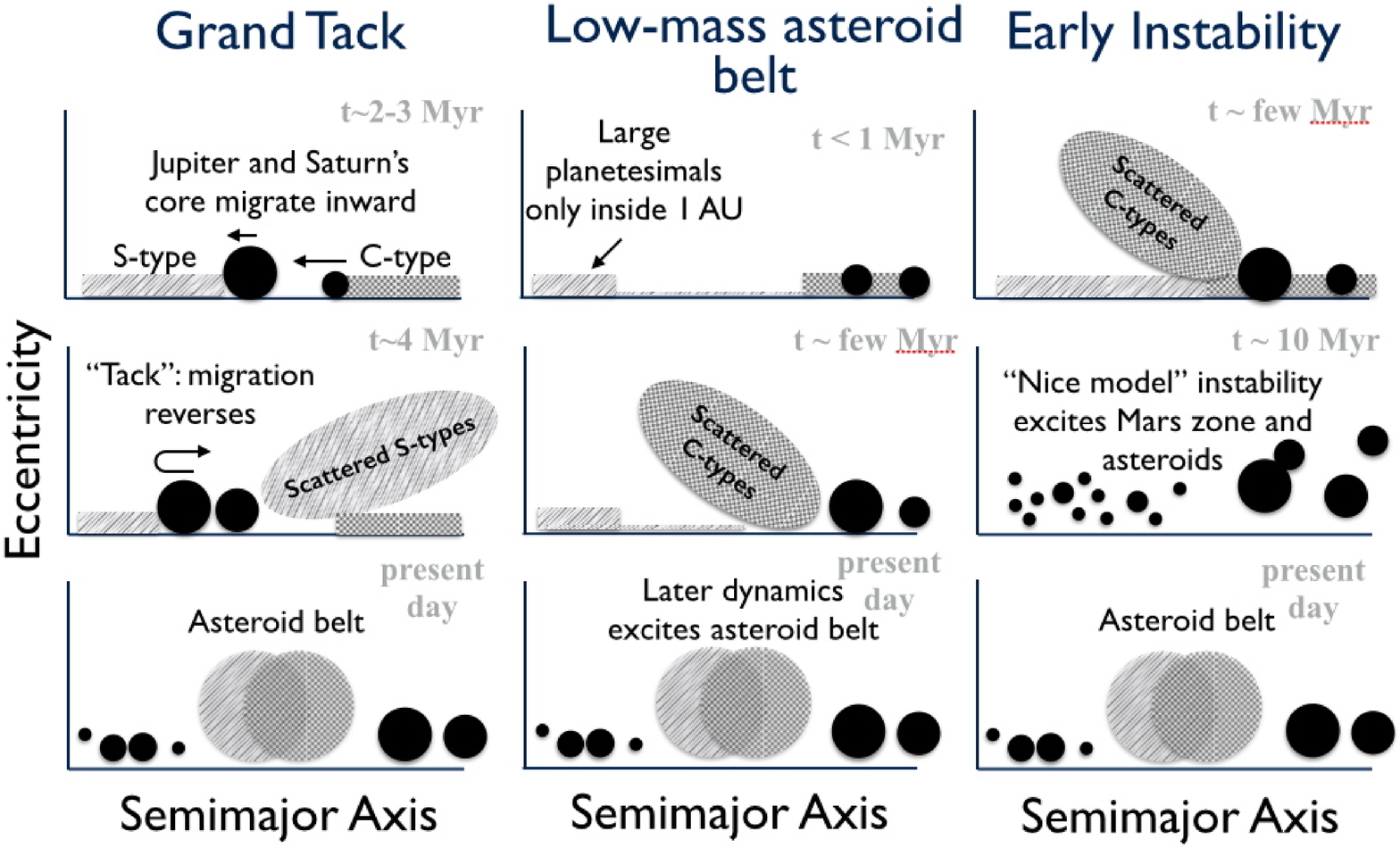} 
    \caption[]{Cartoon of three current global models for Solar System formation, each of which invokes different processes to explain the asteroid belt.  In the {\em Grand Tack} model the asteroid belt is depleted by Jupiter's migration, then re-populated by objects from different regions~\citep{walsh11,walsh12,obrien14}. In the {\em Low-mass asteroid belt} model~\citep[e.g.][]{hansen09,drazkowska16} the asteroids are implanted from both the terrestrial planet-forming and giant planet regions~\citep{bottke06,raymond17,raymond17b}.  In the {\em Early Instability} model~\citep{clement18,clement19b}, the belt is depleted and excited by the giant planets' instability. Adapted from \cite{raymond18d}. }
         \label{fig:global}
         \end{center}
\end{figure}

We now discuss how the dynamical mechanisms described in Sections 2 and 3 fit in a larger context.  Different combinations of processes form the basis of three global models that each broadly match the Solar System: the {\em Low-mass Asteroid belt}, {\em Grand Tack}, and {\em Early Instability} models~\citep[see Fig.~\ref{fig:global} and extensive discussion in][]{raymond18d}. These models are not mutually exclusive. For example, a Grand Tack-like migration of Jupiter and Saturn could have preceded an early giant planet instability. Likewise, a pre-depleted asteroid belt would be consistent with an early instability.  

In Sect. 4.1 we discuss how different combinations of processes can match the asteroid belt constraints laid out in Sect. 1.1, how they compare with global models of Solar System formation, and open questions.  To conclude we discuss the likely evolutionary paths of Vesta and Ceres (Sect. 4.2).  

\subsection{Mechanisms that match the asteroid belt constraints}

Let us consider combinations of dynamical processes that can explain the observed asteroid belt. Based on current thinking, there are three plausible assumptions for the asteroid belt's initial mass distribution. The first is that the belt started with an Earth-mass or more in planetesimals. The second is that the belt was born low in mass, with perhaps a few percent of an Earth-mass in planetesimals. The third is that no planetesimals ever formed within the belt.  In this subsection we discuss how the present-day belt could have been sculpted starting from each of these starting assumptions. We will focus on three constraints: the asteroid belt's very low mass, its orbital excitation, and its compositional diversity (see Sect. 1.1). 

A high-mass primordial asteroid belt requires a depletion of at least three orders of magnitude to match the current one. The asteroids likely started out on dynamically cold orbits. Self-stirring by planetary embryos within the belt (in the framework of the classical model of terrestrial planet formation) has been shown to deplete the belt by roughly two orders of magnitude (see Sect. 3.1). However, for self-stirring to match the asteroid eccentricity and inclination distribution requires such a large mass in asteroidal embryos that simulations cannot avoid producing Mars analogs that are far more massive than the real planet (the so-called `small Mars' problem), and often stranding Mars-mass embryos in the belt (see Sect. 3.3). Jupiter's inward-then-outward migration in the Grand Tack model can deplete the belt sufficiently and also explain the belt's compositional and orbital structure (see Sect. 2.2.2). Its main weakness is that it is unclear whether long-range outward migration can be sustained when gas accretion is taken into account. A third mechanism for strongly depleting the asteroid belt is a violent giant planet instability, which may also explain the belt's excitation (Sect. 3.4). In this case, implantation during giant planet growth would be needed to match the belt's compositional structure (see Sect. 2.1). However, it remains to be seen whether a strong enough instability to deplete the belt by 3 orders of magnitude can remain consistent with the Solar System as a whole. 

A low-mass primordial asteroid belt would only need to be depleted by an order of magnitude to match the current one.  Such depletion is a natural consequence of a relatively gentle giant planet instability (Sect. 3.4). The terrestrial planets' masses and orbits are reproduced if they accreted from a narrow annulus initially located between the orbits of Venus and Earth~\citep{hansen09,kaib15,walsh16,raymond17b,clement19b,deienno19}. The C-types would have been implanted during the giant planets' growth (Sect. 2.1) or migration (Sect. 2.2). 
A Grand Tack-like migration of Jupiter and Saturn (Sect. 2.2.2) or purely inward migration of the giant planets is consistent with a low-mass primordial belt (Sect. 2.2.3). Some planetesimals from the terrestrial region would still be implanted (Sect. 3.2), which would especially be needed in a Grand Tack scenario given Jupiter's strong depletion of asteroidal planetesimals during migration. The asteroids' orbits could have been excited by chaotic phase in the giant planets' orbits (Sect. 3.3) or possibly during the giant planet instability (Sect. 3.4.2). 

An empty primordial asteroid belt must be dynamically populated rather than depleted. The C-types would naturally be implanted from the Jupiter-Saturn region and beyond during the giant planets' growth (Sect. 2.1) and migration (Sect. 2.2). The D-types would have been implanted during the giant planet instability (Sect. 3.4.1), regardless of the belt's initial mass. As for a low-mass asteroid belt, the terrestrial planets are reproduced if they grew from a narrow annulus. Although implantation of planetesimals from the terrestrial region into the belt is efficient enough to explain the S-types (Sect. 3.2), it remains to be seen whether this process is consistent with cosmochemical constraints. The asteroid orbits -- in particular the C-types' -- could have been excited by the giant planets during a chaotic phase (Sect. 3.3) or the giant planet instability (Sect. 3.4.2). 

Moving forward, there remain three central uncertainties for models of the asteroid belt's origin and dynamical evolution. First, the initial mass distribution of the belt in large planetesimals remains. Studies of planetesimal formation in evolving disks~\citep[e.g.][]{drazkowska16,carrera17,drazkowska18} and of the interpretation of dust rings in planet-forming disks~\citep[e.g.][]{alma15,andrews16,andrews18} will address this in the coming years.  Two other uncertainties are related to the giant planets' orbital evolution. The degree to which -- and in what direction -- the giant planets and their cores migrated has a strong impact on the early depletion or implantation and excitation of the belt (see Sect. 2.2). This issue will continue to be addressed with hydrodynamical simulations that include realistic prescriptions for gas accretion. Finally, simulations of the giant planet instability show a spectrum of outcomes (Sect. 3.4.2); limiting which pathways are plausible will constrain the belt's excitation and depletion, and feed back on the belt's starting mass. 

Future constraints will come from continually-improving numerical simulations and observations.  Upcoming large surveys such as LSST will improve the size and quality of asteroid data. Some of the most helpful constraints are those that are not anticipated.  For example, \cite{dermott18} showed that the entire inner asteroid belt may come from just a handful of parent bodies. What this represents in terms of the asteroids' origin and dynamical evolution is not immediately clear. 

\subsection{The dynamical histories of Ceres and Vesta}

We now turn our attention to Ceres and Vesta, the targets of the {\tt Dawn} spacecraft~\citep{russell16}. They are the two most massive asteroids and between them comprise more than 40\% of the belt's total mass.  While Ceres and Vesta currently occupy relatively close orbits, it is likely that they formed in vastly different environments in different parts of the Sun's protoplanetary disk.

Ceres has long been thought to be rich in water ice~\citep[e.g.][]{lebofsky81,fanale89,mccord05}.  {\tt Dawn} confirmed this view and detected a number of volatile species including water ice~\citep[e.g.][]{kuppers14,nathues15,bland16,prettyman17,desanctis17}. Carbonaceous chondrites appear to be the closest meteorite analogs to Ceres~\citep[despite some differences -- ][; see also Chapters 7 and 8]{milliken09,bland16,mcsween18,marchi19}. The {\tt Dawn} spacecraft also detected ammoniated phyllosilicates, are thought to have an outer Solar System origin~\citep{desanctis15}. This evidence suggests that Ceres may been implanted by the same dynamical processes that are thought to have implanted the C-type asteroids~\citep[to our knowledge, the idea that Ceres was implanted from the outer Solar System was first proposed by ][]{mckinnon08}. During the giant planets' gas accretion, planetesimals from the Jupiter-Saturn region and beyond were scattered and implanted into the outer parts of the belt, on orbits that match the C-types' (see Sect. 2.1). The distribution of implanted Ceres-sized planetesimals has a broad peak at $\sim 2.7$~au corresponding to  roughly half of Jupiter's orbital radius~\citep{raymond17}. This peak is a result of the dynamics of scattering by Jupiter. As a result of weak gas drag felt by large planetesimals their orbits are slow to decouple from Jupiter's (by a drop in eccentricity and therefore aphelion) and so they are scattered repeatedly. This creates a much broader distribution in the orbital eccentricities and semimajor axes of large planetesimals, which is eroded from the outside-in as objects with Jupiter-crossing orbits are eventually ejected. This creates a broad peak at half of Jupiter's orbital distance (at the time of scattering), close to Ceres' orbital position. Migration may also have played a role; implantation into the outer main belt is a common outcome in the both the Grand Tack model~\citep{walsh11,walsh12} and in scenarios that invoke inward migration of the giant planets~\citep{raymond17,pirani19}. 

Vesta is a large ($D = 525$~km) differentiated asteroid in the inner main belt whose surface is extensively cratered~\citep{marchi12}. Vesta's Oxygen isotopic composition does not fit within a clear Solar System gradient~\citep[$\Delta^{17}$O $\approx$ -240 ppm, compared with zero and 320 ppm for Earth and Mars, respectively; e.g.][]{clayton96,franchi99,scott02,zhang19}. This argues that Vesta may have been scattered onto its current orbit from closer to the Sun. \cite{bottke06} proposed that Vesta was scattered out from the terrestrial planet-forming region. Multiple studies have shown that this is dynamically plausible~\citep[][see also Sect. 3.2]{bottke06,raymond17b,mastrobuonobattisti17}. Future studies of Vesta's detailed composition may further constrain where in the planet-forming disk it could have accreted~\citep[e.g.][]{toplis13}.

While their present-day orbits cross, Ceres and Vesta likely formed many astronomical units apart. The Jupiter-Saturn region represents the source for most C-types and therefore Ceres' most likely formation region~\citep{mckinnon08,walsh11,raymond17}. If Vesta was implanted from closer-in then its most likely formation distance is exterior to Earth's orbit, perhaps at $\sim 1.5$~au, because the implantation rate is higher for objects originating near Mars' orbit than Earth's~\citep{bottke06,raymond17b}.  In addition, the collisional environment at that distance was less violent than at Earth's orbit~\citep{bottke06}.

On a dynamical note, Ceres and Vesta's have a non-negligible probability of collision in the future~\citep[a probability of roughly 0.2\% per Gyr; ][]{laskar11}. In fact, orbital chaos induced by close encounters between Ceres and Vesta is the limiting factor in predicting the long-term future evolution of the terrestrial planets' orbits. Thus, in addition to providing insights on our Solar System's early history, Ceres and Vesta prevent us from seeing our distant future.

\vskip .4in
{\it Acknowledgments.} 
We thank Rogerio Deienno for a careful review that improved this chapter, and the Editors for several important comments. We are grateful to a long list of colleagues and collaborators including Andre Izidoro, Matt Clement, Alessandro Morbidelli, Seth Jacobson, Nate Kaib, Rogerio Deienno, Bill Bottke, Fernando Roig, Arnaud Pierens, Kevin Walsh, Marco Delbo, Julie Castillo-Rogez, Franck Selsis, Karen Meech, Simone Marchi, Hal Levison, Luke Dones, and David Vokrouhlicky.  S.~N.~R. thanks the CNRS's PNP program and the Agence Nationale pour la Recherche (grant ANR-13-BS05-0003-002 -- {\em MOJO}) for funding and support. D.~N.'s work was supported by NASA's SSW program. Fig.~1. was prepared with data downloaded from mp3c.oca.eu.

\newpage
\bibliographystyle{icarus}
\bibliography{refs}

\end{document}